\documentclass[]{tPHM2e}

\newcommand{\urusi}{URu$_2$Si$_2$}

\begin{document}

\doi{10.1080/14786435.2014.916425}

\issn{}
\issnp{}
\jmonth{10 April}
\jvol{} \jnum{} \jyear{2014}

\markboth{Taylor \& Francis and I.T. Consultant}{Philosophical Magazine}

\articletype{}

\title{Spin-orbit density wave: A new phase of matter applicable to the hidden order state of URu$_2$Si$_2$}

\author{Tanmoy Das$^{\ast}$\thanks{$^\ast$Email: tnmydas@gmail.com \vspace{6pt}}
\\\vspace{6pt} {\em{Theoretical Division, Los Alamos National Laboratory, Los Alamos, New Mexico, 87545, USA}}\\
\vspace{6pt}\received{Received: August 2013} }

\maketitle

\begin{abstract}
We provide a brief review and detailed analysis of the spin-orbit density wave (SODW), proposed as a possible explanation to the `hidden order' phase of URu$_2$Si$_2$. Due to the interplay between inter-orbital Coulomb interaction and spin-orbit coupling (SOC) in this compound, the SODW is shown to arise from Fermi surface nesting instability between two spin-orbit split bands. An effective low-energy Hamiltonian including single-particle SOC and two-particle SODW is derived, while numerical results are calculated by using density-functional theory (DFT) based band structure input. Computed gapped quasiparticle spectrum, entropy loss and spin-excitation spectrum are in detailed agreement with experiments. Interestingly, despite the fact that SODW governs dynamical spin-excitations, the static magnetic moment is calculated to be {\it zero}, owing to the time-reversal invariance imposed by SOC. As a consequence, SODW can be destroyed by finite magnetic field even at zero temperature. Our estimation of the location of the quantum critical point is close to the experimental value of $B_c\sim$~35~T. Finally, we extend the idea of SODW to other SOC systems including iridium oxides (iridates) and two-dimensional electronic systems such as BiAg$_2$ surface and LaAlO$_3$/SrTiO$_3$ interface. We show hints of quasiparticle gapping, reduction of preexisting magnetic moment, large magneto-resistance etc. in these systems which can be explained consistently within the SODW theory.
\bigskip
\bigskip

\end{abstract}


\section{Introduction}
A second order phase transition has been observed in the heavy fermion metal {\urusi} at $T_h$=17.5~K via a sharp discontinuous jump in the specific heat with about 24\% entropy loss.\cite{Cv_Palstra} Considering an associated sharp anomaly in the $c-$axis magnetic susceptibility, it was initially assumed to be an antiferromagnetic (AF) phase.\cite{Cv_Palstra,nonlin_sus2} However, in subsequent works,\cite{Broholm,AFM_HO} including a very recent one,\cite{PDas} the presence of any static magnetic moment has been eliminated. This apparent `magnetic dichotomy' has remained a smoking gun feature of the phase, known as the `so-called' hidden order (HO) phase of \urusi. Additional key fingerprints of the HO state include a relatively weak intensity peak at the commensurate wavevector ${\bf Q}_0$=(001),\cite{commensurate} and a dispersive collective excitation at an incommensurate vector ${\bf Q}_1$=(1$\pm$0.4,0,0),\cite{INS} as probed by inelastic neutron scattering (INS) measurement.  Furthermore, the INS spectral weight loss at the incommensurate wavevector, as opposed to the commensurate one, can fully account for the entropy loss at the HO transition. Direct spectroscopies including angle-resolved photoemission spectroscopy (ARPES)\cite{ARPES,ARPES_FS,ARPEShybridization,ARPESShen,Durakiewicz} scanning tunneling microscopy (STM)\cite{STMDavis,STMYazdani} and point contact spectroscopy\cite{PCT} have found the presence of a Fermi surface (FS) gapping at the HO state, which is consistent with the bulk measurement of about 40\% loss of quasiparticle weight at this phase.\cite{FSgapped} Finally, a Shubunikov de-Hass (SdH) measurement finds that the quantum oscillation frequency does not change by any significant amount in going from HO phase at ambient pressure to the large moment AF (LMAF) phase at high pressure.\cite{Hassinger} Considering also the high-magnetic field phase diagram of {\urusi} where the HO phase is intertwined with the LMAF and possibly with other unknown phases,\cite{HarrisonQCP,magfield} this SdH data can be interpreted along the same line. Therefore, a feasible solution to the HO problem lies in a theory which can explain the `magnetic dichotomy', while accounting for the large FS gapping and entropy change, among others.

According to the Fermi-liquid theory, the HO phase transition is associated with a symmetry breaking static order parameter. Numerous evidence of anomalies probed by INS\cite{commensurate,INS} and spectroscopies\cite{ARPES,ARPES_FS,Durakiewicz,STMDavis,STMYazdani} at either ${\bf Q}_0$ and/ or at ${\bf Q}_1$ suggest that HO breaks translational symmetry. Recent evidence from ultrasonic measurement\cite{lattice1,latticemaple} (and also from ARPES\cite{Santanderprivate}) points to the presence of a lattice anomaly ($\Gamma_3$-type lattice symmetry breaking) at the HO transition, suggesting the involvement of orbital degrees of freedom, in addition to spin. As mentioned before, no evidence of time-reversal symmetry breaking is reported to date.\cite{PDas}  Interestingly, a recent magneto-torque measurement has reported the presence of an in-plane rotational symmetry breaking in a single crystal sample, which, however, has been placed on hold until reproduced.\cite{Matsuda}

This interesting problem has triggered the development of many novel proposals. While these proposals are mainly targeted to solve the HO state in \urusi, they are also interesting in a general ground. Among them, many multipolar orders, including quadrupoles,\cite{quadrupole1,quadruoctu,quadrupole3} octupoles,\cite{quadruoctu,octupole} hexadecapoles\cite{DMFT,hexadeca} to dotriacontapoles or rank-5 orders\cite{rank5,rank5Ikeda} can be highlighted. However, an interesting group analysis study have shown that such higher-rank order parameters are not allowed in \urusi.\cite{multipole} Among dipole orders conventional and unconventional spin,\cite{SDW,AFM_HO,SDWORB,LMAF} orbital,\cite{JT,SDWORB} and other unconventional orders\cite{spinliquid} have been proposed. Variety of theoretical models aiming to explain the `nematic' behavior of the HO state (if indeed exists) are given in Refs.~\cite{Riseborough,spinnematic,rank5Ikeda,Hastatic}. In a third paradigm, the typical concept of hybridization between light and heavy bands has been extended to be modulated with a specific nesting wavevector without spin,\cite{Sasha} with 1/2 spins in both bands,\cite{Riseborough} and with integer and 1/2 spins in heavy and light bands, respectively\cite{Hastatic}. However, multiple spectroscopic data including STM,\cite{STMDavis,STMYazdani} point-contact tunneling,\cite{PCT} and ARPES\cite{ARPEShybridization} as well as resistivity, specific heat and magnetic susceptibility\cite{Cv_Palstra,nonlin_sus2,rho_Palstra} have demonstrated that the hybridization phenomena is insensitive to the HO temperature, and it starts at a much higher temperature (about 60-100~K). This is supported by good consistency of the ARPES data in the paramagnetic phase with first-principles band structure calculation, with some discrepancies lying mainly along the $k_z$ direction.\cite{Durakiewicz}  More interestingly, only the hybridization theory\cite{Sasha} and the present SODW theory assuming incommensurate nesting can only explain the INS dispersion and the spectral weight loss at this wavevector, while no other theory has succeeded in explaining this crucial experimental signature of the HO phase.

Taken together, a consistent and widespread research activity from both theoretical and experimental perspectives have been channeled in the HO research for three decades, and the recent boost to it has been triggered by the high-quality sample preparation and spectroscopic data. These high-quality data have been successful in narrowing down the possible origin of the HO state, however no consensus about the nature of HO phase has yet been reached. In present author's opinion, the solution is hidden in the `magnetic dichotomy' associated with the HO state. These clues have led to the proposal of the SODW which explains this apparent paradox.

The paper includes a coherent combination of new calculations, a review on the prior works, and also hints of future works on SODW. The first-principles bandstructure calculation and the low-energy modelling presented in Secs.~3, 4, and a major part of the discussions in Sec.~5 are taken from Ref.~\cite{DasHOPRX}. A substantial overlap between Secs. 6, 8 and Ref.~\cite{DasSR} can also be expected. Some parts of the application of SODW in other materials which deal with the ARPES results for two-dimensional electron gas with Rashba-type SOC is taken from Ref.~\cite{Das2DEG}. The rest of the calculations and Figs.~3, 4, 5, 6(e)-(h), 7, 10 are produced here.

\section{Mechanism and predictions of the SODW order parameter}

\begin{figure}[top]
\centering
\rotatebox[origin=c]{0}{\includegraphics[width=0.8\columnwidth]{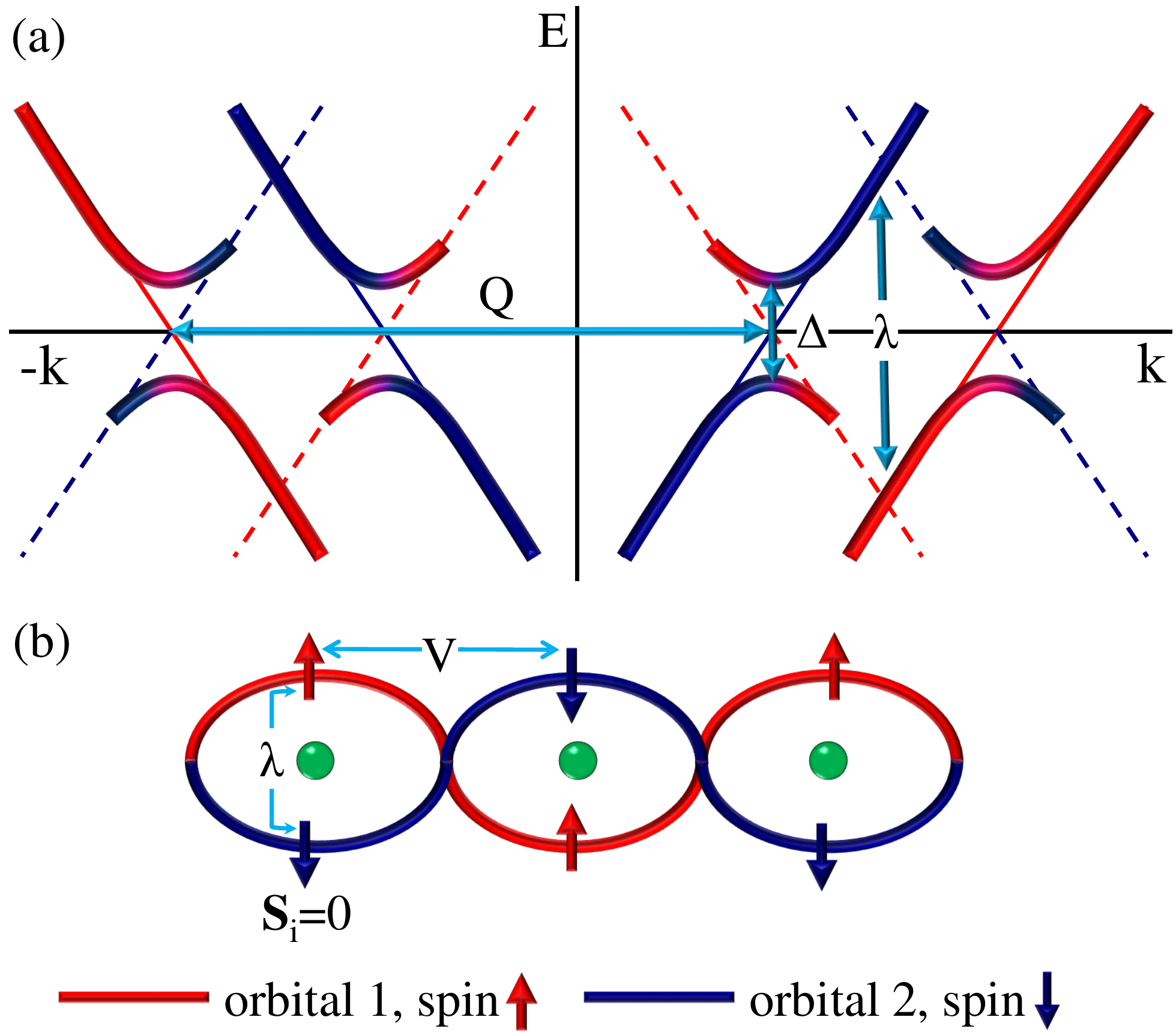}}
\caption{{Schematic illustration of SODW in momentum- and real-space.}  (a) We illustrate the nature of SODW for two SOC split bands in \urusi. Due to the single-particle SOC, each orbital is spin-polarized, but they are mixed in the SODW state determined by the coherence factors. In the case of {\urusi}, two orbital states along (110)-direction in the BCT lattice are nested, and thus becomes gapped at $E_F$.\cite{DasHOPRX} Due to the relevant density wave coherence factors, the quadratic gapped bands on both sides of $E_F$ share different spin-orbital weights, as indicated by red to blue gradient colormap. (b) A real-space view of SODW. Here nearest atomic distance is between two sublattices at which SODW order parameter modulates due to the onsite interaction $V$. But owing to strong atomic SOC, opposite spin polarization in the other orbital state is present (denoted by $\lambda$) at the same site which makes the total spin-moment to vanish at each spatial point, and the Hamiltonian preserves time-reversal symmetry. Of course, perturbations such as crystal distortion or orbital fluctuations may render a finite magnetic moment at each site in this setup, but its magnitude can be expected to be small. Figure taken from Ref.~\cite{DasHOPRX}.}
\label{fig1}
\end{figure}

Recently, it has been shown that SOC can drive new states of matter, including quantum spin-Hall effect,\cite{QSHE} and topological insulator,\cite{TIKane,TIZhang} without electron-electron interaction via symmetry invariance. Another relevant physics of our interest is the theory of Weyl semimetal, originally refereed to `accidental degeneracy in bands'\cite{Weyl}, in which multiple relativistic Dirac points are scattered in the momentum space (and thus termed `semimetal') owing to a broken crystal symmetry, but they remain topologically protected via other subjective symmetry invariance. Considering the `magnetic dichotomy' of the HO phase which apparently breaks time-reversal symmetry (magnetic anomaly) and preserves time-reversal symmetry (no magnetic moment), we introduce the concept of SODW. It breaks translational symmetry by FS nesting, but respects time-reversal invariance owing to the SOC.\cite{DasSR,Das2DEG,DasHOPRX}  While the model is originally applied to \urusi,\cite{DasSR} it has subsequently been extended to two-dimensional electron gas (2DEG) systems in which interface induced Rashba-type SOC is strong, and finite interaction strength can give rise to the SODW and FS gapping,\cite{Das2DEG} consistent with corresponding ARPES data.\cite{ARPES2DEG}

The essential ingredient for the SODW is the presence of (at least) two spin-orbit split bands in the low-energy spectrum, and the FS nesting between them is to be significantly large. If the SOC strength is larger than the Coulomb interaction, an interband spin density wave (SDW) which breaks time-reversal symmetry is disfavored. However, above a critical value of the interorbital Coulomb strength, a SODW can emerge. The schematic view of the SODW is shown in Figs.~\ref{fig1}(a) and \ref{fig1}(b), in momentum and real-space, respectively. In the ${\bf k}$-space, thin solid blue and red lines are two bands, split by SOC $\lambda$, while the corresponding dashed lines are the folded bands shifted by a nesting vector ${\bf Q}$. The SODW order gaps out the `accidental degenerate' points on the FS, and the weight of each orbital is mixed between them. It is evident that SODW can be thought of as two concomitant SDWs with opposite spin state at each spatial point, owing to large SOC. For Rashba-type SOC, the spin-chirality of each band is different to Fig.~\ref{fig1}, and thus the low-energy Hamiltonian changes slightly (see Sec.~\ref{Sec:Other}), however, the general concept is the same.

The consistency of the SODW properties with many experimental signatures of the HO phase in {\urusi} is appealing, particularly as it reconciles the apparently contradictory magnetic properties of this phase. Among them, we highlight several robust features as: (1) The present order breaks translational symmetry,\cite{INS} and spin-rotational symmetry, but thanks to SOC, the order parameter respects time-reversal symmetry and charge conservation at each lattice point. (2) As a result, no magnetic moment is induced, in agreement with measurements.\cite{PDas} (3) Time-reversal symmetry invariance fosters magnetic field to be a perturbation to the order parameter, and the corresponding critical field primarily depends on the gap value and the $g$-factor. Experiments validate the vanishing of the HO around $B\sim35$~T.\cite{magfield}. (4) As a consequence, as field increases the HO gap decreases (given other parameters such as temperature is constant), and thus the resistivity decreases. In other words, a large magnetoresistance effect is expected within the SODW framework. (5) Finally, we offer a detectable prediction of a second spin-1 collective mode localized inside the gap at ${\bf q}\sim 0$ in the SODW state which can be probed via electron spin resonance (ESR) or nuclear quadrupole resonance (NQR) at zero magnetic field or polarized neutron scattering measurements.

Finally, we envisage to take this model further and apply to LaAlO$_3$/SrTiO$_3$ (LAO/STO)\cite{LAOSTO} and iridium-oxide (iridate)\cite{Iridates} where SOC and electronic interaction are large and tunable. In these materials, we present evidence of quasiparticle gapping\cite{Das2DEG,ARPES2DEG}, and strong reduction of preexisting magnetic moment below critical temperature, magnetic field or carrier density, as seen in {\urusi} along the pressure axis.

\section{First-principles band structure and the zone folding in URu$_2$Si$_2$}

\begin{figure}[top]
\centering
\rotatebox[origin=c]{0}{\includegraphics[width=0.8\columnwidth]{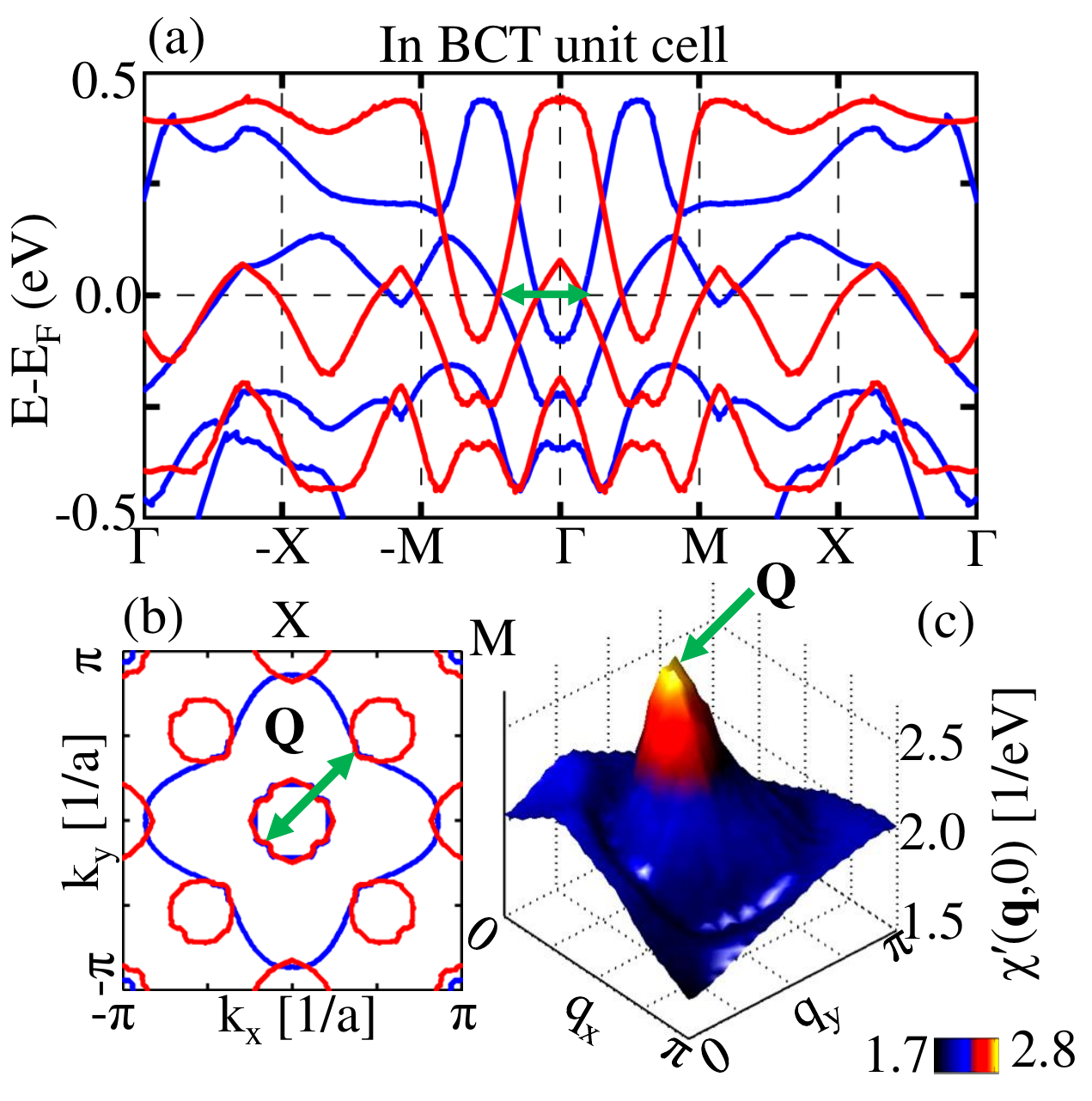}}
\caption{(a) Non-interacting band dispersion (blue line) is plotted along several representative high-symmetry momentum cuts. Red lines are superimposed bands shifted by the nesting vector ${\bf Q}=0.5(\pi, \pi, 0)$ in the BCT unit cell. While a significant region of nesting is visible on the FSs in (b) between original and folded bands along the high-symmetry lines, the dispersions exhibit linear contacts at $E_F$ with Dirac cones. This implies that ${\bf Q}$ nesting causes the development of a density wave in the particle-hole channel with topological properties. (b) FS on $k_z$=0 plane is shown by blue line. Red line is the FS shifted by ${\bf Q}$ vector. Good nesting is evident along the (110)-direction between different orbital states. This diagonal nesting reorients itself along the (100)-direction in the simple tetragonal (ST) lattice.\cite{DasSR} Some additional wiggling in the FS lines arise from finite resolution in the calculation. (c) Static susceptibility is plotted in the 2D {\bf q}-space, affirming a paramount FS instability peak at ${\bf Q}$. Figure extracted from Ref.~\cite{DasHOPRX}.}
\label{fig2}
\end{figure}

In order to determine the symmetry properties of the low-lying states, we begin with investigating the {\em ab-initio} `parent' band dispersion and the FS of {\urusi}\cite{Wien,GGA} in Fig.~\ref{fig2}. The electronic structure in the vicinity of the Fermi level ($E_F$) ($\pm$0.2eV) is dominated by the 5$f$ states of U atom in the entire Brillouin zone.\cite{LMAF,Oppeneer2,STMDavis,Durakiewicz} Due to the lack of the `so-called' itinerant states of $d$- or $p$-characters within the $\pm$500~meV vicinity of $E_F$, the typical hybridization process is weak in this system, and is pushed to a higher temperature $\sim$60-80~K \cite{Cv_Palstra,nonlin_sus2,ARPEShybridization,ARPESShen,STMDavis,STMYazdani}. Furthermore, pertaining to larger atomic size of 5$f$-electrons in U-atoms, than in the largely localized smaller 4$f$-atoms, the inter-atomic electron hopping is larger here and mobile 5$f$ electrons stem within the single-electron picture. In other words, the electronic structure of this actinide compound can be described adequately by first-principles calculation, without invoking Kondo-like physics. In fact, the good consistency between the DFT calculation\cite{LMAF,Oppeneer2,DasSR,DasHOPRX} and ARPES data\cite{Durakiewicz} has excluded the presence of $d$-electrons within the $\pm$500~meV vicinity of $E_F$ to cause any significant hybridization.

Owing to the SOC and the tetragonal symmetry, the 5$f$ states split into the octet $J$=$\frac{7}{2}$ ($\Gamma_8$) states and the sextet $J$=$\frac{5}{2}$ ($\Gamma_6$) states. {\urusi} follows a typical band progression in which the $\Gamma_8$ bands are pushed upward to the empty states while the $\Gamma_6$ states drop to the vicinity of $E_F$. The DFT band structure, calculated by Wien2k software,\cite{Wien,GGA} is shown by blue solid line in Fig.~\ref{fig2}(a) along several representative high-symmetry momentum directions in the body-centered tetragonal (BCT) lattice of the crystal. The corresponding FS topology is depicted in Fig.~\ref{fig2}(b), overlayed by a shifted FS with the `hot-spot' wavevector ${\bf Q}=(\pi/2,\pi/2,0)$ (red lines) in this unit cell notation. Paramount FS nesting is evident here between the two spin-orbit split bands, and hence identified it to be responsible for the SODW state. The FS nesting property is confirmed by static susceptibility calculation, plotted in the basal plane momentum space, in Fig.~\ref{fig2}(c), exhibiting a dominant peak at this ${\bf Q}$ vector. The band structure shifted by ${\bf Q}$ is then plotted in Fig.~\ref{fig2}(a) in red color which reveals that the low-energy bands are linear in momentum, and the contact points between main- and folded-bands at $E_F$ are in the particle-hole channel along the (110) direction. Therefore, the nesting condition in this system resembles the so-called `accidental degenerate points' or Weyl-type Dirac cones\cite{Weyl}.

Since the nesting is between different spin-orbit split bands, a typical SDW is prohibited by time-reversal symmetry and the strong SOC in this case. On the other hand, a spin-orbit entangled order parameter in the two-particle channel can collectively propagate with alternating sign in the total angular momentum at the wavelength determined by the modulation vector. This is the guiding instability that drives spontaneous rotational symmetry breaking, while the time-reversal symmetry remains intact. This is because, both $SU(2)$ groups for spin and orbital separately are odd under time-reversal, but their product $SU(2)\otimes SU(2)$ is even. As the parent state is not a non-trivial topological phase, a gap is opened to lift the FS instability.

\section{Minimal Hamiltonian for the SODW}

Motivated by the above-mentioned experimental results and band structure symmetry properties, we write down the effective theory for the SODW using two SOC split bands. We restrict our discussion to the low-lying $\Gamma_6$ bands and neglect the unfilled $\Gamma_8$ bands. Due to the $j$-$j$-type SOC and time-reversal symmetry, the $\Gamma_6$ atomic states consist of three doublets, characterized by up and down `pseudospins': $m_J$=$\pm$$\frac{5}{2}$,~$\pm$$\frac{3}{2}$,~$\pm$$\frac{1}{2}$, where $m_J$ is the $z$ component of $J$. Although we consider all relevant bands that cross $E_F$ directly from the DFT results for the numerical calculations, however, to explicate the physical mechanism of SODW, and the role of SOC in it, we start from a two-orbital basis with atomistic SOC, $\lambda$. If no other symmetry is broken, the other band (mostly of $J$=$\frac{5}{2}$ character) remains ungapped and thus not considered in the discussion. We, however, note that a nesting between $J$=$\frac{5}{2}$ and $J$=$\frac{3}{2}$ states as reported in Ref.~\cite{Oppeneer2}, can also lead to SODW at a different interaction strength. The purpose here is to derive a general Hamiltonian of such kind, in which no particular assumption of the orbital symmetry and the relevant details of the SOC are imposed. Our starting noninteracting Hamiltonian is thus
\begin{equation}
H_0=\sum_{j{\bf k},\sigma=-{\bar\sigma}}\left[\xi_{j\bf k}c^{\dag}_{j{\bf k},\sigma}c_{j{\bf k},\sigma} + i\sigma\lambda c^{\dag}_{j{\bf k},\sigma}c_{j{\bf k},{\bar{\sigma}}}\right],
\label{H0}
\end{equation}
where $c^{\dag}_{j{\bf k}\sigma}$ is the electron creation operator in the $j^{th}$ orbital with spin $\sigma=\pm$, and $\xi_{j\bf k}$ is its corresponding dispersion spectrum in a translational invariant crystal of Bloch momentum ${\bf k}$. With a translational symmetry breaking at the reduced reciprocal wavevector or the ``hot-spot'' vector ${\bf Q}$, the new spin-orbit basis in the Nambu-space yields $\Psi_{\bf k}= (c_{1{\bf k},\uparrow},~c_{2{\bf k},\downarrow},~c_{1{\bf k}+{\bf Q},\uparrow},~c^{\dag}_{2{\bf k}+{\bf Q},\downarrow})$. The spin-polarized Nambu operator is an important consideration which distinguishes the present SODW order parameter from an inter-orbital SDW, and promotes a {\it zero} magnetic moment as will be evaluated rigorously below. In this basis, we express our interacting Hamiltonian in a matrix form\cite{DasHOPRX} as
\begin{eqnarray}
H_{\bf k}=\left(
\begin{array}{cccc}
\xi_{1{\bf k}} & i\lambda  & 0 & \Delta \\
-i\lambda & \xi_{2{\bf k}} & \Delta^* & 0\\
0        &  \Delta   & \xi_{1{\bf k}+{\bf Q}} & -i\lambda \\
\Delta^* &  0        & i\lambda & \xi_{2{\bf k}+{\bf Q}}\\
\end{array}
\right).
\label{Hint}
\end{eqnarray}
The SODW field parameter $\Delta$ is taken to be complex for generality and will be defined below. This SODW Hamiltonian (\ref{Hint}) is very similar to the one proposed for the Rashba-type SOC system,(see Sec.~\ref{Sec:Other}) and also for 5$d$ system iridates.\cite{DasIridates} Also, it is interesting to compare the Hamiltonian with that for the topological insulator (Eq.~1 of Ref.~\cite{SCZNP}) in which we replace the two-particle gap term $\Delta$ by their hybridization term.

We diagonalize the above Hamiltonian in two steps by the Bogolyubov method. First, we define the Bogolyubov spinor for the SOC term as
\begin{eqnarray}
\left(
\begin{array}{c}
b_{1{\bf k},\sigma}\\
b_{2{\bf k},\bar{\sigma}}\\
\end{array}
\right)=\left(
\begin{array}{cc}
\alpha_{\bf k} & \sigma\beta_{\bf k} \\
-\sigma\beta_{\bf k} & \alpha_{\bf k} \\
\end{array}
\right)\left(
\begin{array}{c}
c_{1{\bf k},\sigma} \\
c_{2{\bf k},\bar{\sigma}}\\
\end{array}
\right),
\label{evec1}
\end{eqnarray}
After this basis transformation, the Hamiltonian in Eq.~\ref{H0} becomes diagonal with eigenstates $E_{1,2{\bf k},\sigma}=\xi^+_{\bf k}\pm E_{0{\bf k}}$, and the corresponding spin-orbit helical weight is $\alpha^2_{\bf k} (\beta^2_{\bf k})=\frac{1}{2}\left(1\pm\frac{\xi^-_{\bf k}}{E_{0\bf k}}\right)$, where $\xi^{\pm}_{\bf k}=(\xi_{1\bf k}\pm\xi_{2\bf k})/2$, and $E^2_{0{\bf k}}=(\xi^-_{\bf k})^2+\lambda^2$.  The subsequent diagonalization onto the Nambu-basis on the reduced Brillouin zone follows similarly, and the spin and orbital notations are retained explicitly. Here the Bogolyubov basis transformation is
\begin{eqnarray}
\left(
\begin{array}{c}
d_{n{\bf k},\sigma} \\
d_{m{\bf k}+{\bf Q},\bar{\sigma}}\\
\end{array}
\right)=\left(
\begin{array}{cc}
u_{\bf k} & \tau_{nm}\sigma v_{\bf k} \\
-\tau_{nm}\sigma v_{\bf k} & u_{\bf k} \\
\end{array}
\right)\left(
\begin{array}{c}
b_{n{\bf k},\sigma} \\
b_{m{\bf k}+{\bf Q},\bar{\sigma}}\\
\end{array}
\right),
\label{evec2}
\end{eqnarray}
where $n=1,2\ne m$ are the band indices, and $\tau_{12}=-\tau_{21}=1$ is the spin-orbit helicity index which changes sign when orbitals are interchanged. Here the canonical density operators $u_{{\bf k}} (v_{{\bf k}})$ take equivalent forms as before, and the corresponding quasiparticle states are replaced by $\tilde{\xi}^{\pm}_{\bf k}=(E_{1\bf k}\pm E_{{2\bf k}+{\bf Q}})/2$, and  $\tilde{E}^2_{0{\bf k}}=(\tilde{\xi}^-_{\bf k})^2+|\Delta|^2$. A schematic view of the SODW split bands and the shared orbital character, determined by coherence factors $u_{\bf k}/v_{\bf k}$ between them is given in Fig.~\ref{fig1}(a). It is interesting to notice that SOC allows the many-particle wavefunction to have a dynamical spin flip on the same orbital at the same momentum as indicated by vertical arrow, which governs a collective ${\bf S}=1$ mode as derived below.

It is worthwhile to combine the above two steps to visualize how the four-vector SODW Nambu operator transforms under translational symmetry breaking while retaining time-reversal symmetry:
\begin{eqnarray}
\left(\begin{array}{c}
c_{1{\bf k},\sigma}\\
c_{2{\bf k},\bar{\sigma}}\\
c_{1{\bf k}+{\bf Q},\sigma}\\
c_{2{\bf k}+{\bf Q},\bar{\sigma}}\\
\end{array}\right)
=\left(\begin{array}{cccc}
~\alpha_{\bf k}u_{{\bf k}} & -\beta_{\bf k}u_{{\bf k}} & ~\bar{\sigma}\beta_{\bf k}v^*_{{\bf k}} & -\sigma\alpha_{\bf k}v_{{\bf k}} \\
~\beta_{\bf k}u_{{\bf k}}  & ~\alpha_{\bf k}u_{{\bf k}} & -\bar{\sigma}\alpha_{\bf k}v^*_{{\bf k}} & -\sigma\beta_{\bf k}v_{{\bf k}} \\
-\sigma\beta_{\bf k}v^*_{{\bf k}}  & ~\bar{\sigma}\alpha_{\bf k}v_{{\bf k}} & ~\alpha_{\bf k}u_{{\bf k}} & -\beta_{\bf k}u_{{\bf k}} \\
~\sigma\alpha_{\bf k}v^*_{{\bf k}}  & ~\bar{\sigma}\beta_{\bf k}v_{{\bf k}} & ~\beta_{\bf k}u_{{\bf k}} & ~\alpha_{\bf k}u_{{\bf k}} \\
\end{array}
\right)
\left(\begin{array}{c}
d_{1{\bf k},\sigma}\\
d_{2{\bf k},\bar{\sigma}}\\
d_{1{\bf k}+{\bf Q},\sigma}\\
d_{2{\bf k}+{\bf Q},\bar{\sigma}}\\
\end{array}\right).
\label{Unitary}
\end{eqnarray}

An immediate ansatz emerges from the above transformation $-$ guided by the same symmetry of the Hamiltonian in Eq.~\ref{Hint} $-$ that the system is time-reversal invariant under the representation of this symmetry $\mathcal{T}=\mathbb{I}\otimes i\tau_y\mathcal{K}$, where $\mathcal{K}$ is the complex conjugation, $\mathbb{I}$ is the 2$\times$2 identity matrix, and $\tau_y$ is the second Pauli matrix. Under time-reversal, the Hamiltonian transforms as $H_{\bf k}=\mathcal{T}H_{-{\bf k}}\mathcal{T}^{-1}=\mathbb{I}\otimes\tau_yH^*_{-{\bf k}}\mathbb{I}\otimes \tau_y$. This important symmetry consideration renders a {\it zero} magnetic moment as depicted schematically in Fig.~\ref{fig2}b and is evaluated in Eq.~(\ref{magmoment}) below.

\subsection{SODW order parameter}

\begin{figure}[top]
\centering
\rotatebox[origin=c]{0}{\includegraphics[width=0.7\columnwidth]{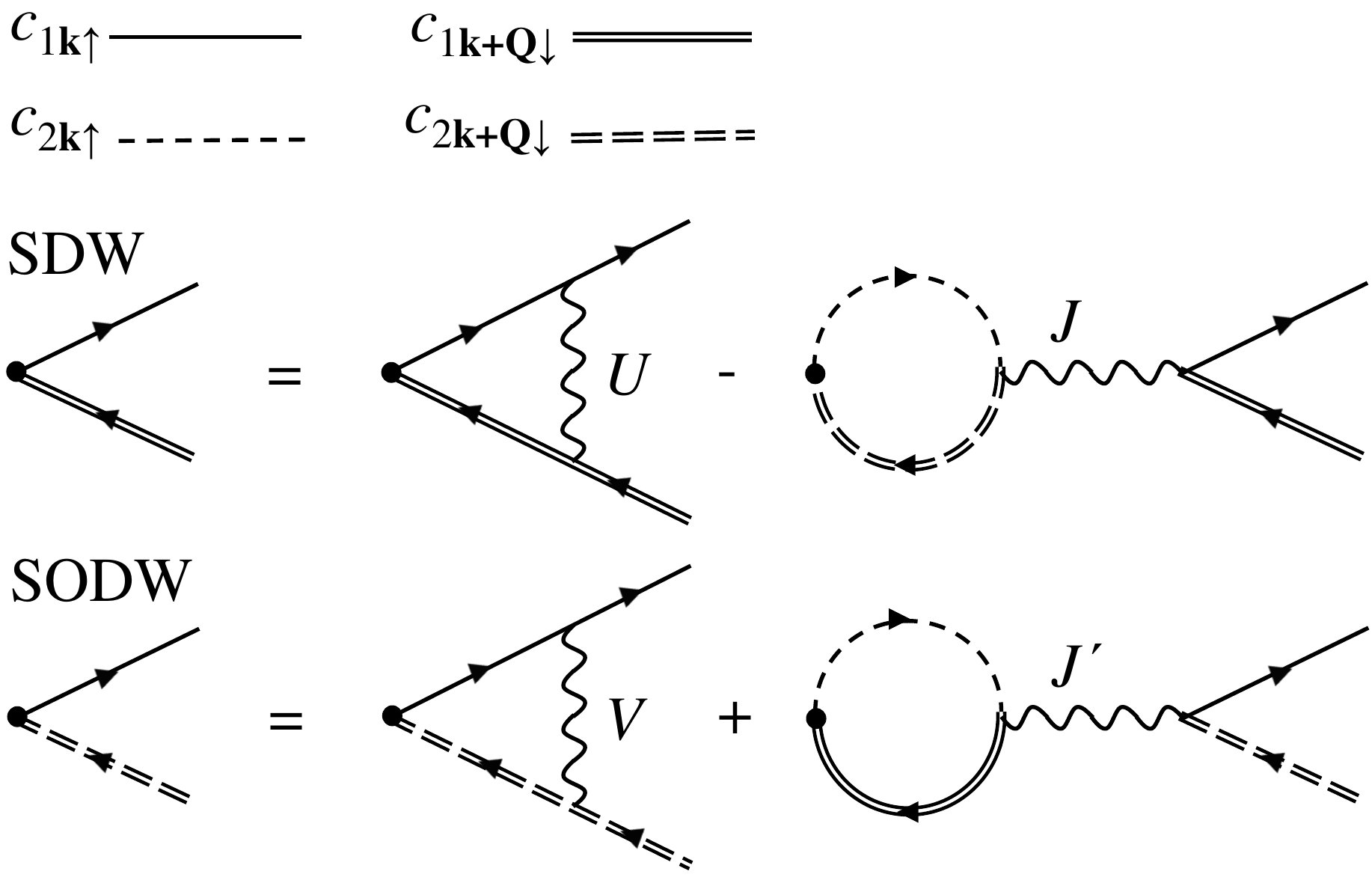}}
\caption{Feynman diagram representations of SDW and SODW in a two spin-polarized orbital systems (having SOC) under translational symmetry breaking perturbation are shown by considering all possible contractions of the interaction Hamiltonian in Eq.~(\ref{RPAHam}). Forward and backward arrows dictate creation and annihilation operators, respectively. Therefore, opposite arrows in each order parameter and bubble indicate that they are in the particle-hole channel.}
\label{fig3}
\end{figure}

\begin{figure}[here]
\centering
\rotatebox[origin=c]{0}{\includegraphics[width=0.5\columnwidth]{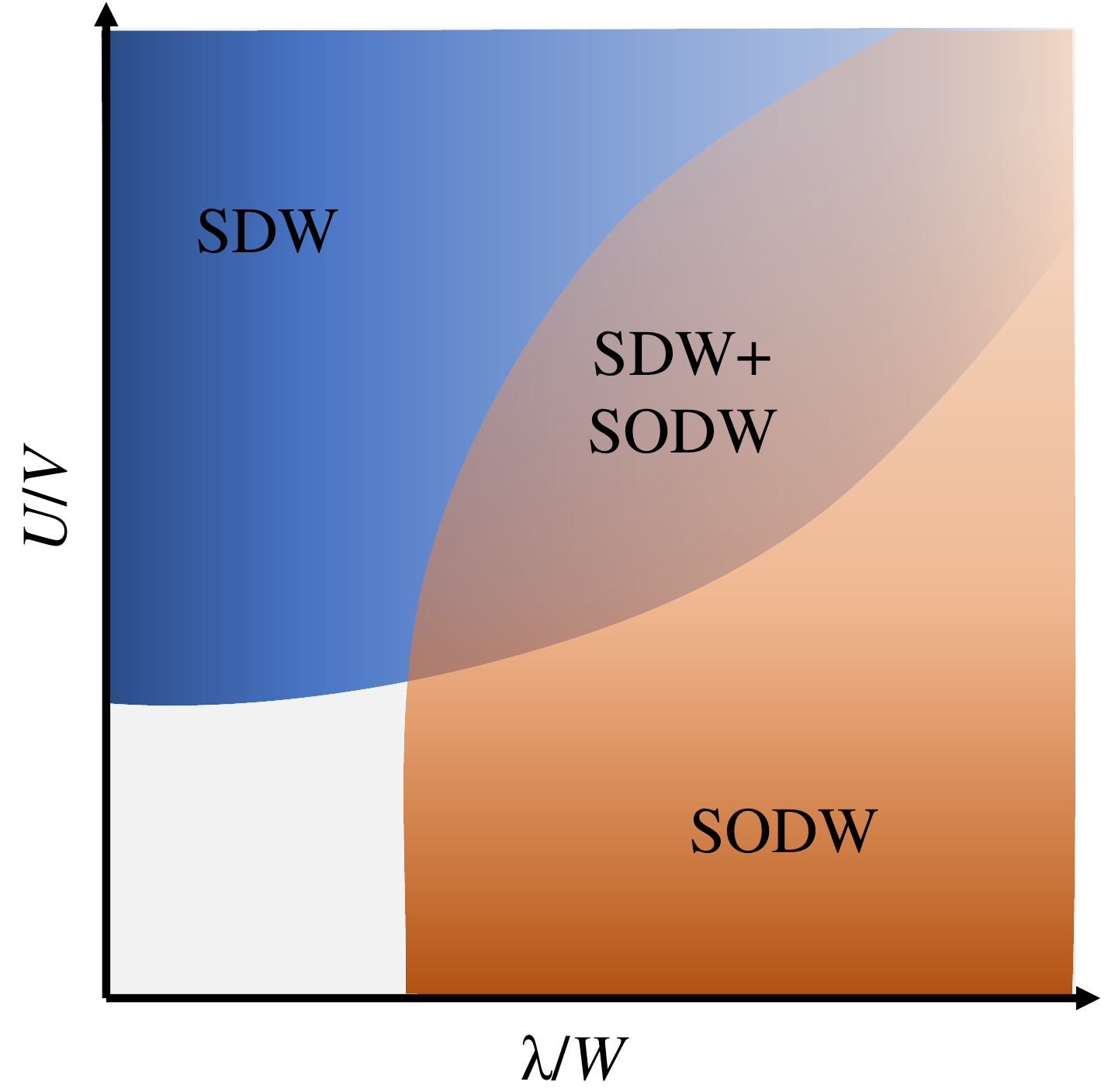}}
\caption{Schematic phase diagram of SDW and SODW phases in the relevant parameter space. Along the horizontal-axis, the strength of SOC $\lambda$ with respect to the corresponding bandwidth ($W$) is plotted, while the vertical-axis gives the ratio between the intraorbital ($U$) and interorbital ($V$) Coulomb interactions. Large values of SOC and $V$ drive SODW, while large $U$ favors SDW, and their delicate balance can lead to a coexistence between them. Other relevant interaction terms such as Hund's coupling and pair scattering which contribute to the SDW and SODW orders, respectively, can be assumed to have similar but smaller effects, see Eq.~(\ref{sceqn}).}
\label{fig4}
\end{figure}

In a multiorbital setup, the full interacting Hamiltonian includes intra- and interorbital Coulomb interactions, $U$ and $V$, Hund's coupling $J$, and pair-exchange term $J^{\prime}$:
\begin{eqnarray}
H_{int}&=&\sum_{{\bf k}_1-{\bf k}_4}\left[\sum_i U c^{\dag}_{i{\bf k}_1,\uparrow}c_{i{\bf k}_2,\uparrow}c^{\dag}_{i{\bf k}_3,\downarrow}c_{i{\bf k}_4,\downarrow}\right.\nonumber\\
&&+\sum_{i<j,\sigma}\left(Vc^{\dag}_{i{\bf k}_1,\sigma}c_{i{\bf k}_2,\sigma}c^{\dag}_{j{\bf k}_3,\bar{\sigma}}c_{j{\bf k}_4,\bar{\sigma}}+(V-J)c^{\dag}_{i{\bf k}_1,\sigma}c_{i{\bf k}_2,\sigma}c^{\dag}_{j{\bf k}_3,\sigma}c_{j{\bf k}_4,\sigma}\right)\nonumber\\
&&\left. +\sum_{i<j,\sigma}\left(Jc^{\dag}_{i{\bf k}_1,\sigma}c^{\dag}_{j{\bf k}_3,\bar{\sigma}}c_{i{\bf k}_2,\bar{\sigma}}c_{j{\bf k}_4,\sigma}
J^{\prime} c^{\dag}_{i{\bf k}_1,\sigma}c^{\dag}_{i{\bf k}_3,\bar{\sigma}}c_{j{\bf k}_2,\bar{\sigma}}c_{j{\bf k}_4,\sigma} + h.c.\right)\right].
\label{RPAHam}
\end{eqnarray}
where ${\bf k}_1+{\bf k}_3={\bf k}_2+{\bf k}_4$. The linearized equations for the two order parameters SDW and SODW are presented graphically in Fig.~\ref{fig4} which leads to the self-consistent equations as $1=-T_{n}\Gamma_n$ where $n$=`SDW', and `SODW'. For a given order, we consider all possible contractions of the interaction term written in Eq.~2, which lead to the following vertices
\begin{eqnarray}
\Gamma^i_{SDW}&=& U\chi_{ii}^{ii}({\bf Q},\omega=0) + J\chi_{ii}^{jj}({\bf Q},\omega=0)~({\rm for}~i=1,2\ne j),\nonumber\\
\Gamma_{SODW}&=& V\chi_{12}^{12}({\bf Q},\omega=0) + J^{\prime}\chi_{12}^{21}({\bf Q},\omega=0).
\label{sceqn}
\end{eqnarray}
The corresponding susceptibilities in the particle-hole channel are then defined as
\begin{eqnarray}
\chi_{il}^{jk}({\bm q},\omega)
&=&-\frac{1}{N}\sum_{{\bm k},\mu\nu}\phi_{\mu}^i({\bf k})\phi_{\mu}^{j\dag}({\bf k})\phi_{\nu}^k({\bf k}+{\bf q})\phi_{\nu}^{l\dag}({\bf k}+{\bf q})\frac{f(E_{\nu}({\bm k}+{\bf q}))-f(E_{\mu}({\bm k}))}{\omega+i0^+ + E_{\nu}({\bm k}+{\bf q})-E_{\mu}({\bm k})}.\nonumber\\
\label{chi}
\end{eqnarray}

It is interesting to see that Hund's coupling does not contribute to the SODW, while the pair scattering term does. However, in many materials, this term has very small value and can be neglected for some practical purposes. The vertices in Eqs.~(\ref{sceqn}) are different here from a typical multiband system (see, for example, Ref.~\cite{Daspnictide}) due to the presence of SOC. Based on Eqs.~\ref{sceqn} and \ref{chi}, we can deduce a general phase diagram in Fig.~\ref{fig4} for the competition and coexistence of SODW and SODW. For a single orbital system, only an intraband nesting $\chi_{ii}^{ii}$ is present and thus only SDW is allowed. When SOC is included, the inter-orbital susceptibility $\chi_{ij}^{ij}$ is turned on. The relative competition between these two nestings determines the dominant instability to win in a given band structure. At the limit of $J=J^{\prime}=0$, when SOC $\lambda\rightarrow 0$, no SODW is possible, similarly, as $U\rightarrow 0$, no SDW is possible from interorbital susceptibility (since $J=0$ is also imposed). For large $V$, i.e. small $U/V$, SDW is more stable for finite $\lambda$. When $U/V\rightarrow 1$ (assuming both intra- and interorbital nesting are comparable), two phases can coexist. From Eq.~\ref{sceqn}, it is also easy to deduce that while inter-orbital susceptibility helps SDW via Hund's coupling $J$, no intra-orbital term is involved in the SODW phase. 

If there is no orbital fluctuation present in the system, for the two fully spin-polarized bands, the magnetic moment along any direction vanishes. This is achievable in materials with strong SOC as in {\urusi} in which $\lambda\sim$1~eV. This can be verified by a trivial analytical computation of the spin-operator:
\begin{eqnarray}
S&=&\frac{1}{N}\sum_{i{\bf k}}\langle c^{\dag}_{i{\bf k}+{\bf Q},\alpha}\sigma^i_{\alpha\beta}c_{i{\bf k},\beta}\rangle\nonumber\\
&=&\frac{1}{N}\sum_{i{\bf k}}\alpha_{\bf k}\beta_{\bf k}\left[\sigma\left(-u_{\bf k}v^*_{\bf k}+u_{\bf k}v^*_{\bf k}\right)\langle d^{\dag}_{1{\bf k},\sigma}d_{1{\bf k},\sigma}\rangle \right.\nonumber\\
&&\hspace{1.9cm}\left.+ \bar{\sigma}\left(-u_{\bf k}v_{\bf k}+u_{\bf k}v_{\bf k}\right)\langle d^{\dag}_{2{\bf k},\bar{\sigma}}d_{2{\bf k},\bar{\sigma}}\rangle + \left({\bf k}\rightarrow {\bf k}+{\bf Q}\right)\right] = 0.
\label{magmoment}
\end{eqnarray}
Magnetic moment vanishes in each band for any value of the interaction strength, owing to the presence of two opposite SDW coupled by SOC. We should carefully distinguish the induced magnetic moment of SODW in Eq.~({\ref{magmoment}}), from a SDW terms (a finite gap term to be placed in the `13' component of Hamiltonian in Eq.~(\ref{Hint}) which has of course different interaction term as discussed in Eq.~\ref{sceqn}). Our conclusion of zero moment is only valid when no SDW term is explicitly present.

Finally, the SODW order parameter can then be written in an explicit form as
\begin{eqnarray}
\sigma \Delta_0&=&\frac{V}{N}\sum_{ij{\bf k}}\left\langle  c^{\dag}_{i{\bf k}+{\bf Q},\alpha}i\tau^y_{ij}\sigma^x_{\alpha\beta}c_{j{\bf k},\beta}\right\rangle
=\frac{2V}{N}\sum_{\bf k}^{\prime} u_{\bf k}{\rm Re}\left[v_{\bf k}\right]\left[f(\tilde{E}^+_{\bf k})-f(\tilde{E}^-_{\bf k})\right],
\label{SODW}
\end{eqnarray}
where the fermion occupation number is defined as $f(\tilde{E}^+_{\bf k})=\langle d^{\dag}_{1{\bf k},\sigma}d_{1{\bf k},\sigma}\rangle$ and so on, and $\tilde{E}^{\pm}_{\bf k}$ are the degenerate eigenstates of the interaction Hamiltonian in Eq.~\ref{Hint}. The notation `prime' over the momentum summation indicates that the summation is restricted within the reduced Brillouin zone. The complex gap parameter that enters in the Hamiltonian is $\Delta=\Delta_0(\sigma_x+\sigma_y)$.  For the experimental value of gap amplitude $\Delta_0\sim 10$~meV, we estimate the critical inter-orbital interaction strength to be $V\sim0.6$~eV which is a reasonable number for 5$f$ electrons estimated earlier.\cite{DasPRL,DasPRX}

\subsection{Competition and coexistence of HO and LMAF phases}\label{Sec:HOLMAF}

\begin{figure}[here]
\centering
\rotatebox[origin=c]{0}{\includegraphics[width=0.99\columnwidth]{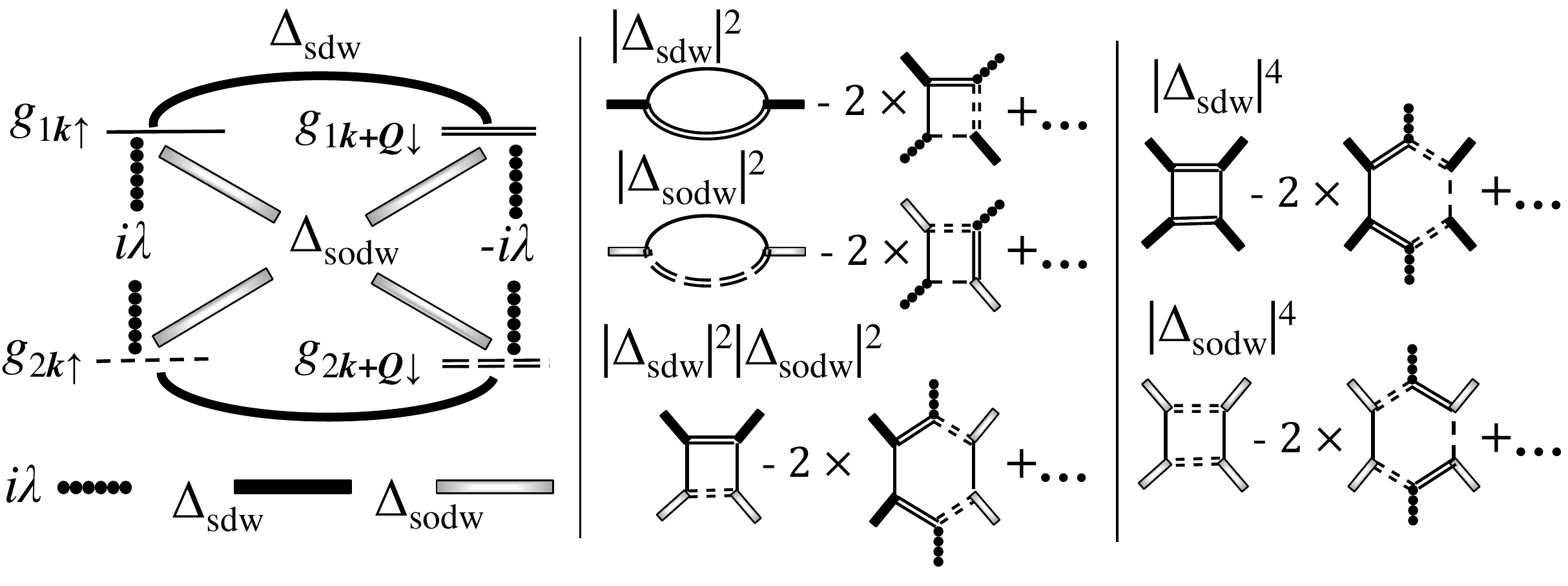}}
\caption{Feynman diagram of various GL coefficients for SODW and SDW order parameters. The green's function $g_{i{\bf k}\sigma}$ and $g_{i{\bf k}+{\bf Q}\sigma}$ are assumed to be for particle and hole states, respectively, or vice versa.}
\label{fig5}
\end{figure}

A fascinating property of the HO state in {\urusi} is that it transforms into the LMAF phase via a first-order phase transition as pressure (and also magnetic field) is increased. SdH oscillation measurement\cite{Hassinger} has demonstrated that the angle-dependence of the oscillation frequency, which is taken to be proportional to the closed FS areas at finite magnetic field, does not change as we go from the HO phase to the LMAF phase via increasing pressure. This observation suggests that both HO and LMAF phases coexist in {\urusi}, at least at high magnetic field. Based on this result, we study the stability of the two phases, attributing the LMAF phase as a SDW state, within a Ginzburg-Landau (GL) treatment by integrating out the diagonal and the SOC terms (single-particle terms). Here we denote $\Delta_{\rm sodw}$, and $\Delta_{\rm sdw}$ for the SODW and SDW order parameters. The GL free energy can be expressed as
\begin{eqnarray}
\mathcal{F}[\Delta_{\rm sodw},\Delta_{\rm sdw}] = -T\sum_{{\bf k},n}{\rm ln}~{\rm det}G^{-1}_{{\bf k},n} + \frac{|\Delta_{\rm sodw}|^2}{V} + \frac{|\Delta_{\rm sdw}|^2}{U},\nonumber\\
\label{GLF0}
\end{eqnarray}
where the inverse Green's function in the Nambu-Grassmann space is ${\hat G}^{-1}_{{\bf k},n}=i\omega_n\mathbb{I}-H_{\bf k}$, with $i\omega_n$ is the Matsubara frequency for the fermions, and $\mathbb{I}_{4\times 4}$ is the identity matrix. Expanding Eq.~\ref{GLF0} up to the fourth order in the gap fields, we obtain the essential GL functional for a homogeneous system of real fields as
\begin{eqnarray}
\mathcal{F}[\Delta_{\rm sodw},\Delta_{\rm sdw}] = \sum_{i={\rm sdw, sodw}}\left[\alpha_i|\Delta_i|^2+\beta_i|\Delta_i|^4\right]+ \gamma|\Delta_{\rm sodw}|^2|\Delta_{\rm sdw}|^2.
\label{GLF}
\end{eqnarray}
All expansion coefficients $\alpha_i$, $\beta_i$, and $\gamma$ are shown in Fig.~\ref{fig5} which lead to
\begin{eqnarray}
\alpha_1&=& \frac{1}{U} - \chi^{(2)}_{1} + 2\lambda^2\chi^{(4)}_1,\\
\alpha_2&=& \frac{1}{V} - \chi^{(2)}_{2} + 2\lambda^2\chi^{(4)}_2,\\
\beta_i&=& -\chi^{(4)}_i + 2\lambda^2\chi^{(6)}_{i},\\
\gamma&=& -\chi^{(4)}_3 + 2\lambda^2\chi^{(6)}_{3}.
\end{eqnarray}
For simplicity the SDW order parameter for both states are assumed to be the same. The two, four and six order susceptibilities $\chi^{(2),(4),(6)}$ are defined in the band basis as
\begin{eqnarray}
\chi^{(2)}_{1} &=& T\sum_{{\bf k},n} g_{1{\bf k},n}g_{1{\bf k}+{\bf Q},n},\\
\chi^{(2)}_{2} &=& T\sum_{{\bf k},n} g_{1{\bf k},n}g_{2{\bf k}+{\bf Q},n},\\
\chi^{(4)}_{1} &=& T\sum_{{\bf k},n} (g_{1{\bf k},n}g_{1{\bf k}+{\bf Q},n})^2,\\
\chi^{(4)}_{2} &=& T\sum_{{\bf k},n} (g_{1{\bf k},n}g_{2{\bf k}+{\bf Q},n})^2,\\
\chi^{(4)}_{3} &=& T\sum_{{\bf k},n} g_{1{\bf k},n}g_{1{\bf k}+{\bf Q},n}g_{1{\bf k},n}g_{2{\bf k}+{\bf Q},n},\\
\chi^{(6)}_{1} &=& T\sum_{{\bf k},n} g_{1{\bf k},n}g_{1{\bf k}+{\bf Q},n}g_{2{\bf k}+{\bf Q},n}g_{2{\bf k},n}g_{2{\bf k}+{\bf Q},n}g_{1{\bf k}+{\bf Q},n},\\
\chi^{(6)}_{2} &=& T\sum_{{\bf k},n} g_{1{\bf k},n}g_{2{\bf k}+{\bf Q},n}g_{1{\bf k}+{\bf Q},n}g_{2{\bf k},n}g_{1{\bf k}+{\bf Q},n}g_{2{\bf k}+{\bf Q},n},\\
\chi^{(6)}_{3} &=& T\sum_{{\bf k},n} g_{1{\bf k},n}g_{1{\bf k}+{\bf Q},n}g_{2{\bf k}+{\bf Q},n}g_{1{\bf k},n}g_{2{\bf k}+{\bf Q},n}g_{1{\bf k}+{\bf Q},n}.\\
\end{eqnarray}
Here the Green's functions $g$ denote the diagonal components of the full Green's function matrix as $g_{j{\bf k},n}=(i\omega_n-\xi_{j{\bf k}})^{-1}$. The phase transition for each order parameter is determined as usual by the corresponding quadratic terms as $\alpha_{i}=\alpha_{0i}(T-T_i)$, where $T_i$ are the transition temperatures for them without phase competition. For multiple competing order parameters, a general criterion was derived before in various contexts\cite{Rafael,chubukov,DasGL} that both phases can share a uniform coexistence region if $0<\gamma^2<\beta_1\beta_2$. Whereas for $\gamma^2>\beta_1\beta_2$, the phase fluctuation is sufficiently strong that both phases are separated by a first-order phase transition. Given the experimental fact in {\urusi} that HO and LMAF phases share a first-order phase boundary as a function of pressure, the second criterion is expected in this material. However, the numerical calculations of the susceptibilities and the effects of pressure are beyond the scope of the present paper.

\section{Electronic and magnetic anomalies of SODW}

We give two representative results of the electronic and magnetic structure of the SODW in the HO state and compare them with ARPES and INS data, respectively.

\subsection{Quasiparticle gapping and entropy count}
\begin{figure}[top]
\rotatebox[origin=c]{0}{\includegraphics[width=1.0\columnwidth]{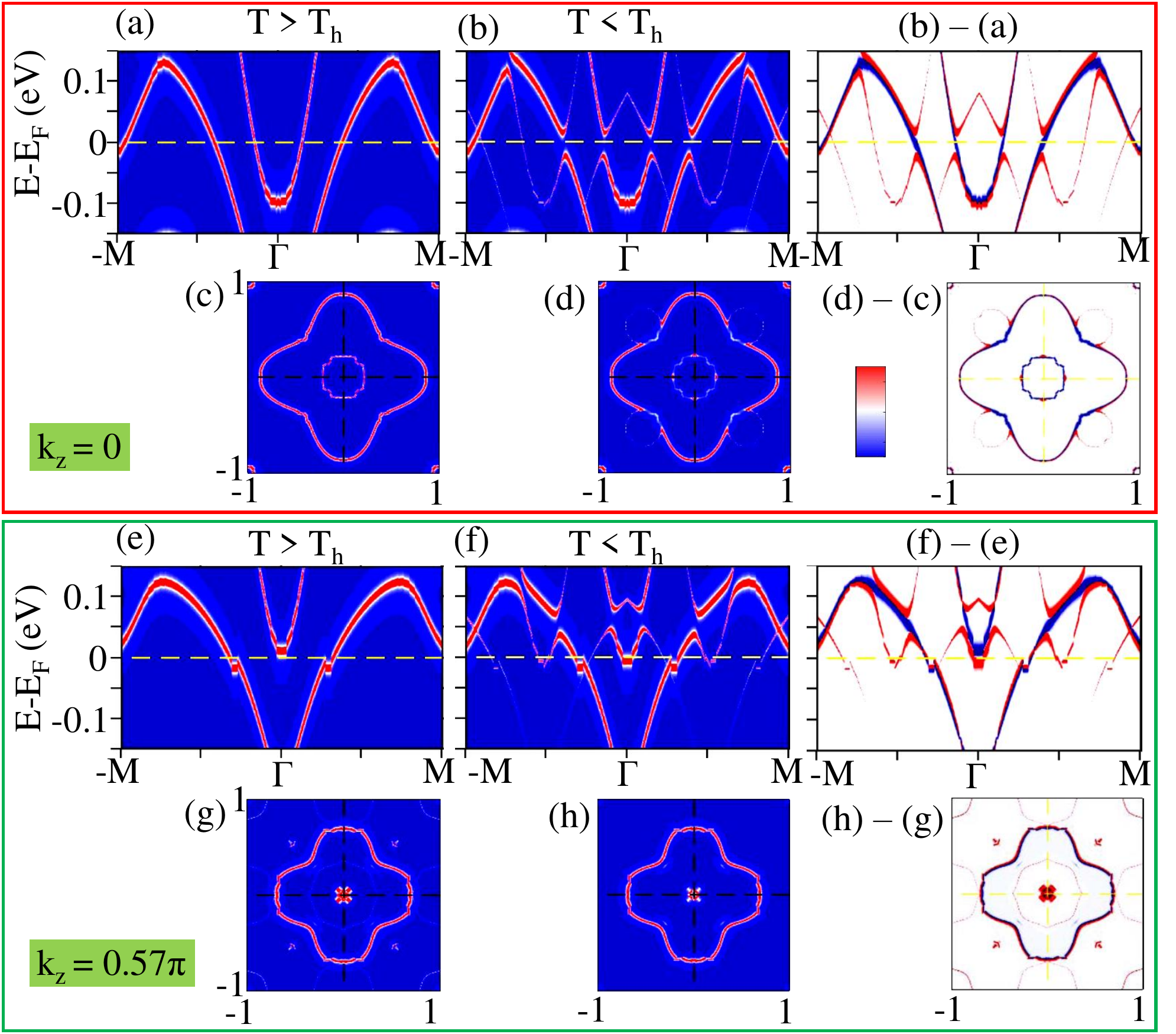}}
\caption{{Images of electronic fingerprints of the SODW state, and the `surprising' collapse of conduction band (at $k_z = 0.57\pi$) at the HO transition.} (a) Spectral weight maps of the bare band dispersion plotted along the (110) direction above the HO transition, deduced from DFT calculation with a constant broadening of 0.5~meV. (b) Corresponding quasiparticle dispersion is shown in the HO state within the SODW scenario. Clear semiconductor-like gap at $E_F$ is observed for the two bands centering the $\Gamma$-point (indicated by arrow). (b)-(a) The difference in spectral weight between the HO and paramagnetic states is shown here. The difference highlights the shift of band around M-point as pointed out in an ARPES data (Fig.~3 of Ref.~\cite{Durakiewicz}) (c), (d) Computed FS below and above the HO transition, respectively, plotted at the $k_z=0$ plane. A good correspondence between theory and ARPES result (Fig.~2 of Ref.~\cite{Durakiewicz}) can be noted here. (c)-(d) The corresponding spectral weight differences are shown  at $E_F$. (e)-(f) Computed spectral function of dispersion on a characteristic $k_z=0.57\pi$ plane in the paramagnetic and HO states, respectively. The `unusual' collapse of the conduction band below $E_F$ is reproduced in theory due to the opening of the HO in the empty state on this $k_z$-plane. This intriguing result is observed in ARPES data\cite{ARPES} along the same momentum cut at an incident photon energy of 21.2~eV in Ref.~\cite{ARPES}, in reasonable agreement with the $k_z$ value in theory. (g), (h) Corresponding FS on the same $k_z$-cut confirms the development of a tiny electron pocket centering $\Gamma$-point below the HO transition. We do not find any other $k_z$-cuts, except near $k_z=0$, where the SODW-induced HO parameter gaps out most of the FS. An artificially large gap value of 20~meV is chosen here for visualization, where the actual gap lies in between 5-10~meV range. The figure for $k_z=0$-plane (top panel) is reproduced from Ref.~\cite{DasHOPRX}. }
\label{fig6}
\end{figure}

We first evaluate the electronic signature of HO gap opening. The SODW order induced HO state is evaluated by using DFT-based band structure as the input frame of reference, and downfolding the full DFT results in the reduced space near $E_F$ with the `hot-spot' wavevector ${\bf Q}=(\pi/2,\pi/2,0)$ in the BCT crystal structure. Figure~\ref{fig6} gives single-particle maps along several representative momentum cuts, and the FS topology before and after the HO transition. Gapping of the FS is clearly visible for the two bands aligned along the (110) direction at the degenerate Dirac points at $E_F$, introduced in Fig.~\ref{fig1}(a). This gapping process truncates the paramagnetic FS into small pockets aligned along the bond-direction. We require to pay particular attention to the location of the FS pockets in this system, since the original crystal structure is BCT (not simple tetragonal which is often used for the simplification of computation). In this unit cell notation, the location of the gapping and FS pockets governed in the SODW state are in direct agreement with the ARPES spectral function map on the FS.\cite{Durakiewicz} We find that the loci of the gapped states move away from $E_F$ as we increase the $k_z$ value.

ARPES has also reported an interesting result that a conduction band collapses below $E_F$ at the $\Gamma$-point upon entering into the HO state.\cite{ARPES} This unusual behavior is also reproduced within the present framework, without invoking the previously-thought hybridization phenomena. As mentioned before, the optimum FS nesting occurs at the $k_z=0$-plane, and beyond it, the same nesting condition moves to empty states as the $\Gamma$-centered electron pocket gradually shrinks. This is the manifestation of the itinerant nature of the HO gap which generates an anisotropic gap on the electronic structure, and can also induce momentum dependence in the $g$-factor measured in quantum oscillation measurements.\cite{dHvA} We locate the $k_z\sim0.57\pi$ plane where the bottom of the paramagnetic electron pocket resides just above $E_F$, and upon the HO gap opening in the empty state, this pocket drops into the Fermi sea. This result, shown in the bottom row of Fig.~\ref{fig6}, agrees well with ARPES measurements.\cite{ARPES} Since ARPES can probe different $k_z$-planes with the variation of incident photon energy, our theoretical result is consistent with the experimental fact that such phenomena is observed at a finite photon energy of $h\nu=$~21.2~eV. Our explanation can further be confirmed by observing the possible flickering presence of the $\Gamma$-centered electron pocket as a function of the photon energy.

\begin{figure}[top]
\centering
\rotatebox[origin=c]{0}{\includegraphics[width=.7\columnwidth]{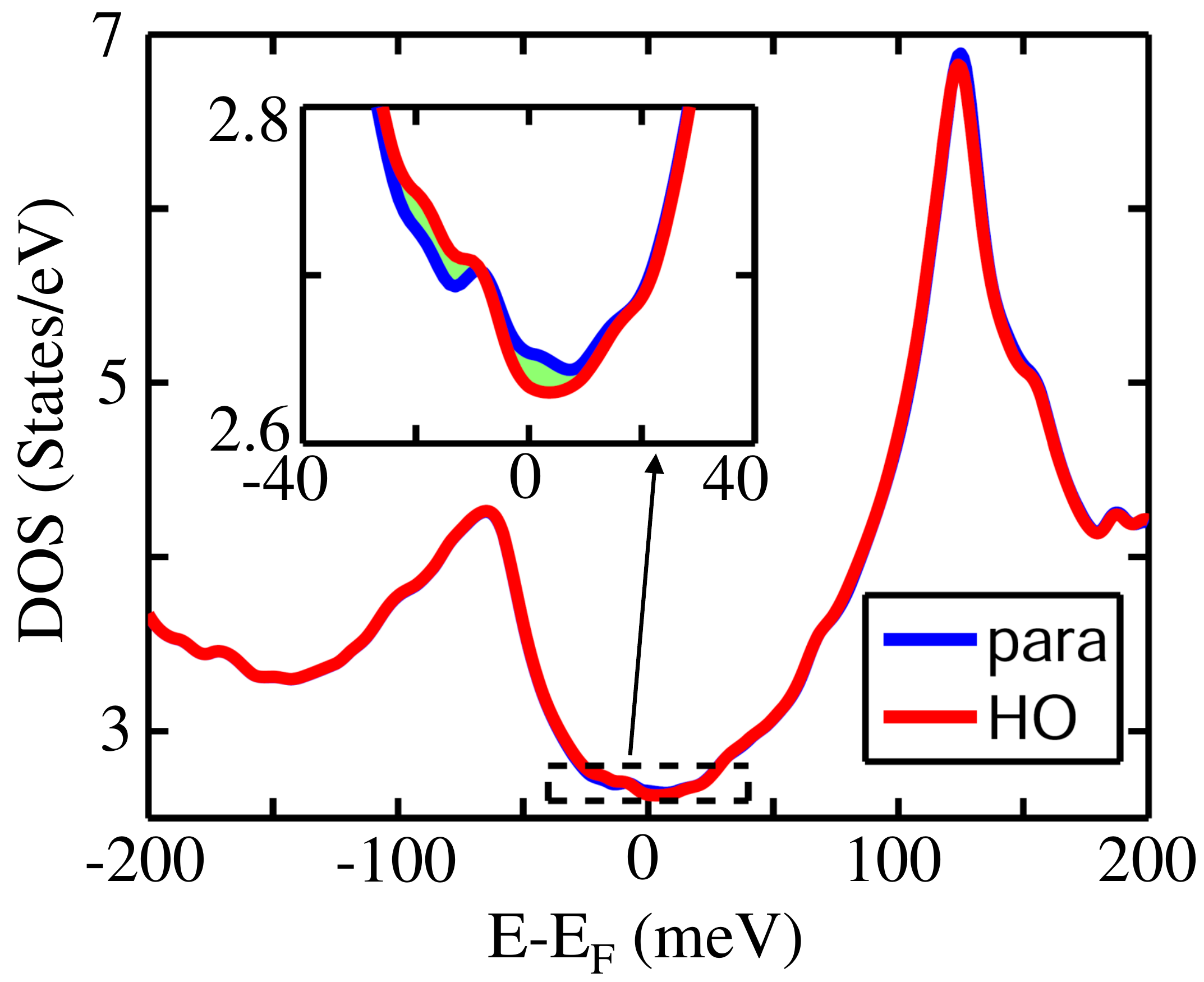}}
\caption{DOS below (red) and above (blue) the HO transition. Two van-Hove singularities or flat band-like features are observed below and above $E_F$ which yield large spin- and pseudospin fluctuation in the particle-hole channel and thus govern large mass enhancement in the electronic states as seen in other actinide compounds.\cite{DasPRL,DasPRX} Near $E_F$ the HO gap opens, and the corresponding residual spectral weight shifts below $E_F$.}
\label{fig7}
\end{figure}

The corresponding density of states (DOS) before and after the HO transition are shown in Fig.~\ref{fig7}. Without invoking any hybridization, we see two flat band-like peaks (or the van-Hove singularities) below and above $E_F$ (which are pushed further towards lower energy by renormalization). The HO gap opens around $E_F$ as seen from the {\it inset} box in Fig.~\ref{fig7} and the corresponding spectral weight shifts to the states at a higher energy below $E_F$. The results are consistent with STM measurements.\cite{STMDavis,STMYazdani}

In this context, we note an interesting consequence of the two flat-bands in the particle-hole channel. As shown in Pu-115 and U-115 families,\cite{DasPRL,DasPRX} such particle-hole channel, triggered by SOC, causes large dynamical spin and pseudospin fluctuations which dress the electronic states in the intermediate coupling regime. A trademark feature of the intermediate coupling scenario is the duality between renormalized itinerant quasiparticle states and localized high energy states in the same or different orbitals, separated by the (pseudo-) spin fluctuation energy scale, which is seen in ARPES.\cite{Kawasaki} The pseudospin fluctuation is also reported in a recent NMR study in this system.\cite{NCurro} This is a future problem to study in \urusi.

The entropy release of the HO transition can be easily evaluated from the Free-energy estimation. The Free-energy of the HO state is deduced from $F=-k_BT{\rm ln}({\rm Tr} e^{-H/k_BT})+N\mu$, where $k_B$ is Boltzmann constant, $N$ is number of filled states, and $\mu$ is the chemical potential. In the diagonal basis with a mean-field order, the Free energy translates into $F=-k_BT\sum^{\prime}_{\bf k} {\rm ln}[\sum_{\nu=\pm}e^{-\tilde{E}_{\bf k}^{\nu}/k_BT}]+N(\mu+|\Delta|^2/V)$. And the corresponding entropy release at the HO transition is evaluated to be $\Delta S=(\partial F/\partial T)_{T_h=17.5~K}\sim0.3k_B{\rm ln}2$, which is close to its experimental estimate from the specific heat jump.\cite{Cv_Palstra}

\subsection{Magnetic excitation spectrum}

\begin{figure}[top]
\rotatebox[origin=c]{0}{\includegraphics[width=.99\columnwidth]{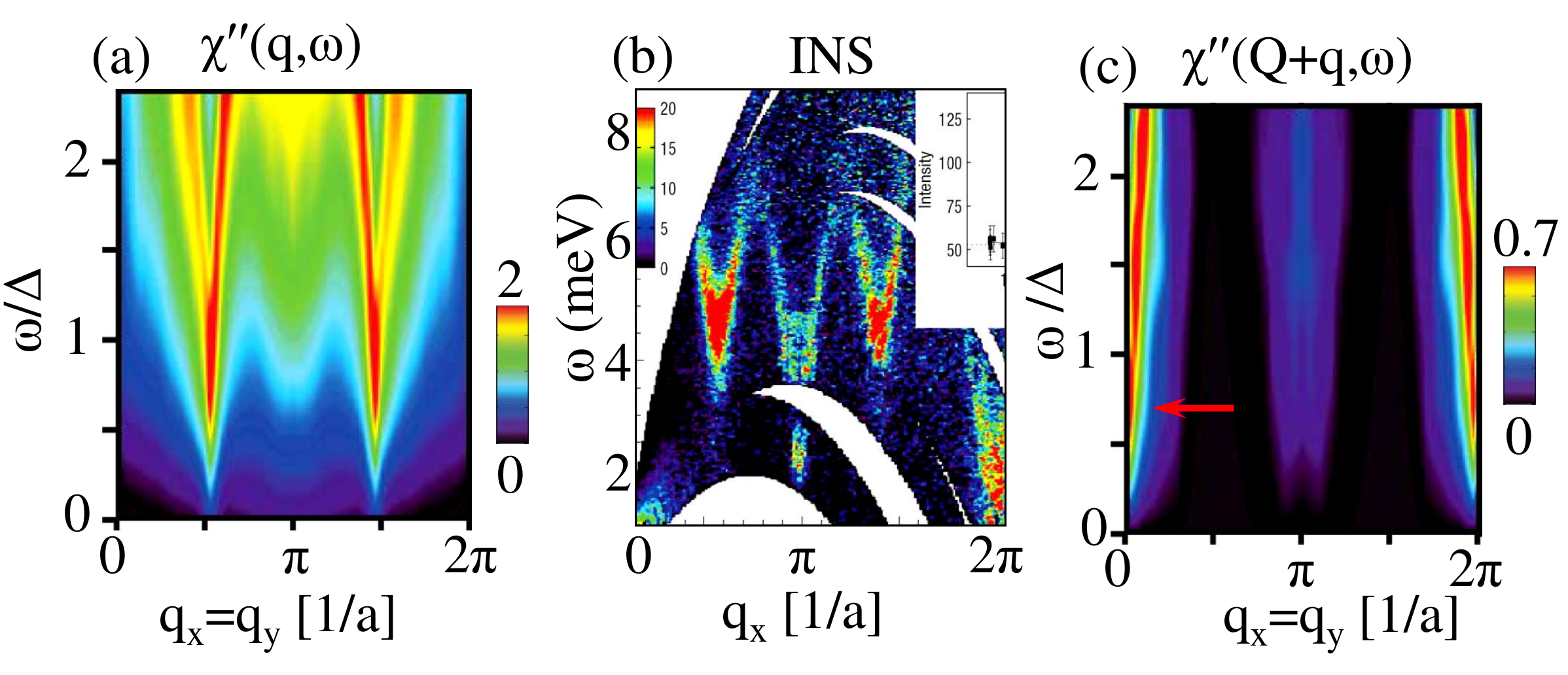}}
\caption{{Computed spin-excitation spectrum and comparison with available INS data.}
Imaginary part of the spin-excitation spectrum plotted along the ${\bf q}$=(110) direction as a function of excitation energy in the HO state. (a) For the inter-band (or equivalently inter-orbital) components, we observe the development of a collective mode inside the HO gap with an upward linear-in-energy dispersion starting from ${\bf Q}\sim 0.55(\pi,\pi,0)$. Although a good correspondence between theory and experiment is visible when compared with the INS data\cite{INS} in (b), a lattice transformation between the BCT to the ST lattice is required to align the (100)-direction in the experimental figure to our plot along the (110) direction. The gapped Goldstone mode or roton-mode is well reproduced via SODW order parameter at this incommensurate wavevector, which deviates slightly from the embedded HO wavevector of ${\bf Q}$ due to band structure effect. (b) INS data in the HO state is taken from Ref.~\cite{INS}. (c) Umklapp spin susceptibility exhibits a second collective mode, localized at ${\bf q}\sim0$ with intensity, which is an order of magnitude smaller than that of the intra-band one. This mode can be probed via ESR or NQR measurements. Figure extracted from Ref.~\cite{DasHOPRX}.}
\label{fig8}
\end{figure}

Now we present the results of the spin-excitation spectrum as a function of both momentum and energy. INS measures the imaginary part of the spin-susceptibility $-$ modulo materials specific form factor $-$ which is theoretically computed from Eq.~(\ref{chi}). In the SODW state, the susceptibility is a 2$\times$2 matrix (to be precise it is a 4$\times$4 matrix including both intra and interband contributions separately as is done in the main computation), made of direct transitions (in the diagonal terms) and Umklapp scattering terms placed in off-diagonals\cite{spinbag}:

\begin{eqnarray}
\chi({\bf q},{\bf q}^{\prime},\omega) =
\left(\begin{array}{cc}
\chi({\bf q},{\bf q},\omega) & \chi({\bf q},{\bf q}+{\bf Q},\omega)\\
\chi({\bf q},{\bf q}+{\bf Q},\omega) & \chi({\bf q}+{\bf Q},{\bf q}+{\bf Q},\omega)\\
\end{array}\right)
\label{chiSODW}
\end{eqnarray}
Here the direct term $\chi({\bf q},{\bf q},\omega)$ means that the scattering between the two main bands and between the two shadow bands while the Umklapp scattering term $\chi({\bf q}+{\bf Q},{\bf q}+{\bf Q},\omega)$ arises from the scattering betwee then main and shadow bands. All scattering processes included here are in the particle-hole channel. Although, the mechanism and structure of the INS results presented in Fig.~\ref{fig7} are tied mainly to  the details of the electronic structure and the HO `hot-spot' wavevector embedded within this bare susceptibility, however, for completeness we invoke the many-body correction within the random-phase approximation (RPA). RPA calculation is justified in this heavy-fermion systems at low-$T$ where itinerant electronic structure is well-established. The RPA interaction Hamiltonian is same as given in Eq.~\ref{RPAHam}, with $J=J^{\prime}=0$. Different bandwidths of different orbitals amount to different critical values of $U$, and $V$, however, to justify that the final results are parameter free, we take the lowest critical value of $U$=1~eV and $V$=0.6~eV which shifts the energy scale of all the excitation mode slightly to a lower energy.

Figure~\ref{fig8} presents the computed direct and Umklapp spin excitation spectra in (a) and (c), respectively, along the zone diagonal direction, and compares with the corresponding experimental data\cite{INS} obtained along the zone boundary direction given in Fig.~\ref{fig8}(b). For direct transition in the particle-hole continuum (trace of the susceptibility matrix), we clearly mark a prominent ${\bf S}=1$ collective mode with a dispersion that resembles gapped Goldstone mode or a roton-like spectrum. For the case of the discrete symmetry breaking due to the complex order parameter $\Delta^*$, such gapping of the Goldstone mode can bee justified. The mode is localized around the ${\bf Q}\sim0.55(\pi,\pi,0)$ value, slightly shifted from the original HO wavevector ${\bf Q}$ due to band structure effect. However, the shift of the mode below $\omega<\Delta$ is a many-body correction as discussed above. To match the mode energy of the experimental data about $\omega=$~5~meV, a gap amplitude of $\Delta_0\sim 6$~meV is invoked in our calculation which is close to the spectroscopic value of the HO gap.\cite{STMDavis,STMYazdani} Although a static neutron signal is reported in Ref.~\cite{PDas} at ${\bf q}=(\pi,\pi,0)$, but in the inelastic spectrum, both experiment\cite{INS} and our theory find a weaker and featureless intensity at this wavevector.

\subsection{${\bf q}\sim 0$ collective mode}

By construction, the SODW is associated with an interaction induced spin-orbit entangled electronic structure, rendering conceptual  similarity with the spin-orbit order in the particle-particle channel proposed by Leggett\cite{Leggett} for the liquid $^3$He superconducting phase, dynamical generation of SOC,\cite{CWu}, or the single-particle quantum spin-Hall state\cite{QSHE} which do not break time-reversal symmetry. In what follows, the Umklapp scattering process between different spin-orbital states brings out two Zeeman-like spin-split states at the same momentum {\bf k}, on both sides of the HO gap with coherence factors $u_{\bf k}/v_{\bf k}$, without any external magnetic field, as discussed in Fig.~\ref{fig1}(a). Therefore, a second spin-flip collective mode is expected in the off-diagonal susceptibility in Eq.~(\ref{chiSODW}) as $\chi({\bf q},{\bf q}+{\bf Q},\omega)\sim \sum_{\bf k}\delta(\omega-\xi_{1{\bf k}}+\xi_{2{\bf k}+{\bf Q}+{\bf q}})$ to localize inside the SODW gap $\omega\le|\Delta|$ around ${\bf q}\sim 0$. Indeed, our computation confirms the existence of this collective mode at energy $\omega/\Delta\sim0.75$, as shown in Fig.~\ref{fig8}(c). Since the intensity of this mode is about an order of magnitude lower than that of the ${\bf Q}$-mode presented in Fig.~\ref{fig8}(a), it will be difficult to simultaneously detect them together in the INS measurement. However, ESR or NQR, having the capability of detecting lineshifts of resonance without the application of a static magnetic field, will be able to measure our proposed ${\bf q}\sim0$ mode.

\section{SODW under magnetic field}

\begin{figure}[top]
\centering
\rotatebox{0}{\scalebox{.65}{\includegraphics{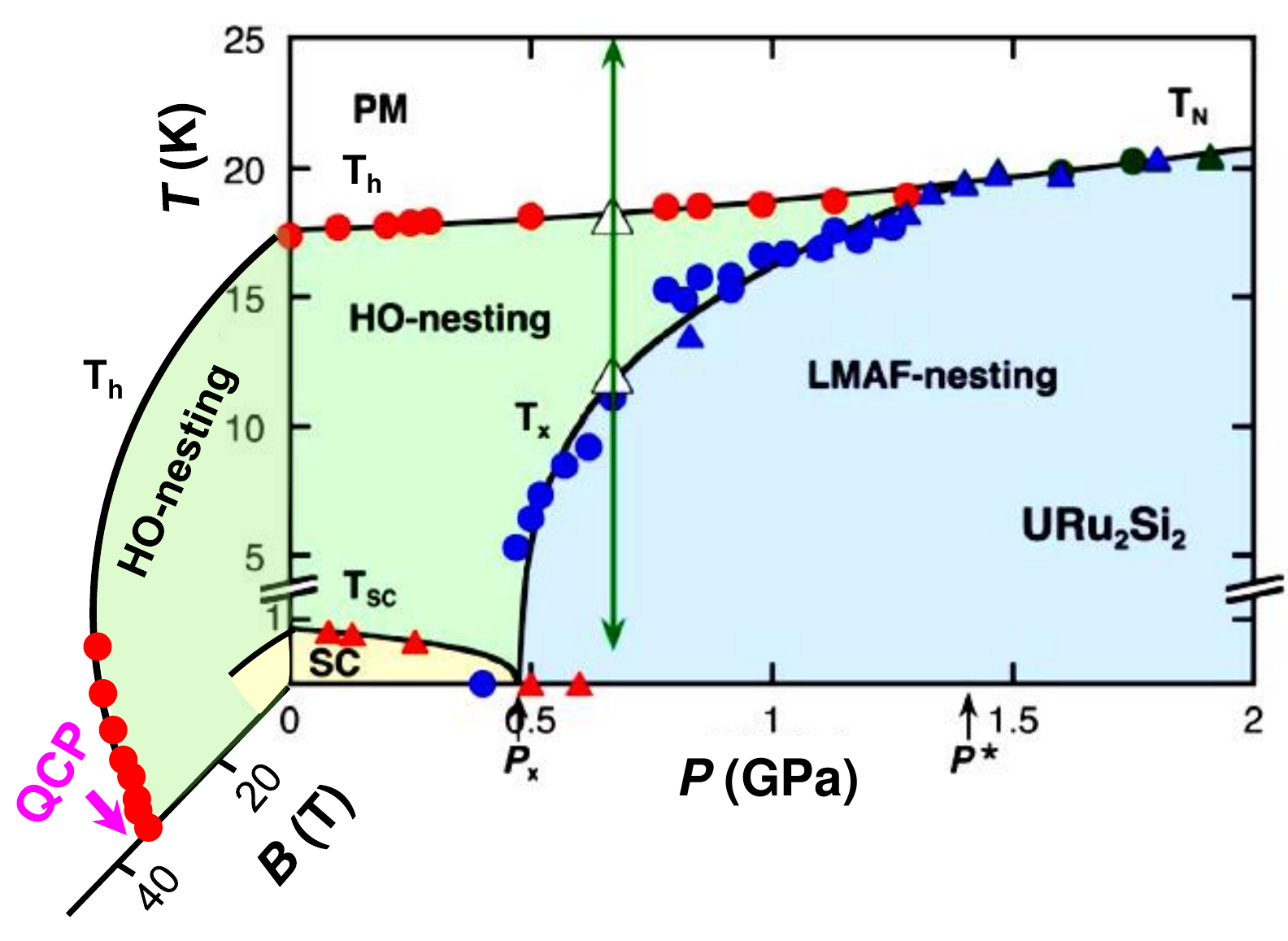}}}
\caption{ Phase diagram of {\urusi} along the magnetic field ($B$) and pressure ($P$) directions. The data on the $P-T$ plane is taken from Ref.~\cite{phase_diagram}. The symbols along $B-T$ plane are extracted from Ref.~\cite{HarrisonQCP} with more similar data available in Ref.~\cite{magfield} (not included in the phase diagram here). A QCP at $T=0$ along the field direction is expected from our theory, and is also observed in experiment.\cite{HarrisonQCP} Although we theoretically overestimate the critical field (see text), we normalize it to match the experimental value for visualization. Deducing the phase diagram for SC and other possible phases that may arise around or above the QCP is beyond the scope of the present calculation. Figure reproduced from the supplementary information of Ref.~\cite{DasSR}}
\label{fig9}
\end{figure}

The present spin-orbit order is not only topologically protected by time-reversal symmetry, but most importantly it is thermodynamically shielded with a condensation energy equal to the quasiparticle gap. The HO gap is insensitive to any time-reversal invariant perturbation such as pressure (with sufficient pressure the HO transforms into the LMAF phase), while time-reversal breaking perturbation such as magnetic field will destroy the order. Remarkably, these are the hallmark features of the HO states,\cite{magfield,HarrisonQCP} which find a natural explanation within our SODW order scenario. In what follows, the magnetic field will destroy the HO state even at $T=0$~K, that means at a quantum critical point (QCP) as the HO is a spontaneously broken symmetry phase. However, due to the finite gap opening at the HO state, it requires finite field value to destroy the order. The thermodynamical critical field can be approximately determined by the critical Zeeman term $E_Z\sim 2g\mu_B B$ ($g$ is `g'-factor, $\mu_B$ is Bohr magneton) which can overcome the condensation gap energy (i.e. $E_z=\Delta$). Assuming a mean-field like magnetic field dependence of the SODW gap $\Delta$, we obtain an approximate phase diagram of the HO state as a function of $B$. The critical field is obtained to be $B_c$ = 43~T for an isotropic $g$-factor of $\sim$1 for 5$f$ electrons and zero field gap of $\Delta=$~5~meV. The result is shown in Fig.~\ref{fig9}. Experimental value of $B_c$ is 35~T.\cite{HarrisonQCP} HO interaction can increase the value of the $g$-factor which will reduce $B_c$. In fact, a SdH oscillation measurement reports the presence of an anisotropic $g$-factor in the HO state,\cite{dHvA} which whether can be reproduced by the SODW order or not is to be explored in a future study.

The present model cannot deduce the phase diagram for the superconducting (SC) state, possibly intervening the HO state, or any other phases that may arise around or above the QCP.\cite{HarrisonQCP} However, from the study of Sec.~\ref{Sec:HOLMAF}, we can expect the appearance of a SDW phase near this phase boundary. Experimentally, it is known that the second order phase transition at low field transforms into a first order one at high field, which may play a role in eliminating the QCP of the SODW state in this system. However, the QCP as a function of magnetic field is a general result for SODW and can thus be searched in other materials.

In this context, we note an interesting result in the LaAlO$_3$/SrTiO$_3$ (LAO/STO) interface.\cite{LAOSTO} This interface is known to host local ferromagnetism and 2DEG state. It is shown in a recent magnetoresistance and anomalous Hall effect experiment that below a critical value of the magnetic field, the magnetic polarization is dramatically reduced while resistivity sharply rises, see Fig.~\ref{fig10}(c). This unique phenomenon indicates the emergence of a new phase which onsets below a critical field and quenches magnetic moment while reduces the electric conductivity via FS gapping. These are the key properties of the SODW. Similar effect of large magnetoresistance is also expected in {\urusi}.

\section{SODW in other spin-orbit coupled systems}\label{Sec:Other}

\begin{figure}[top]
\centering
\rotatebox[origin=c]{0}{\includegraphics[width=.99\columnwidth]{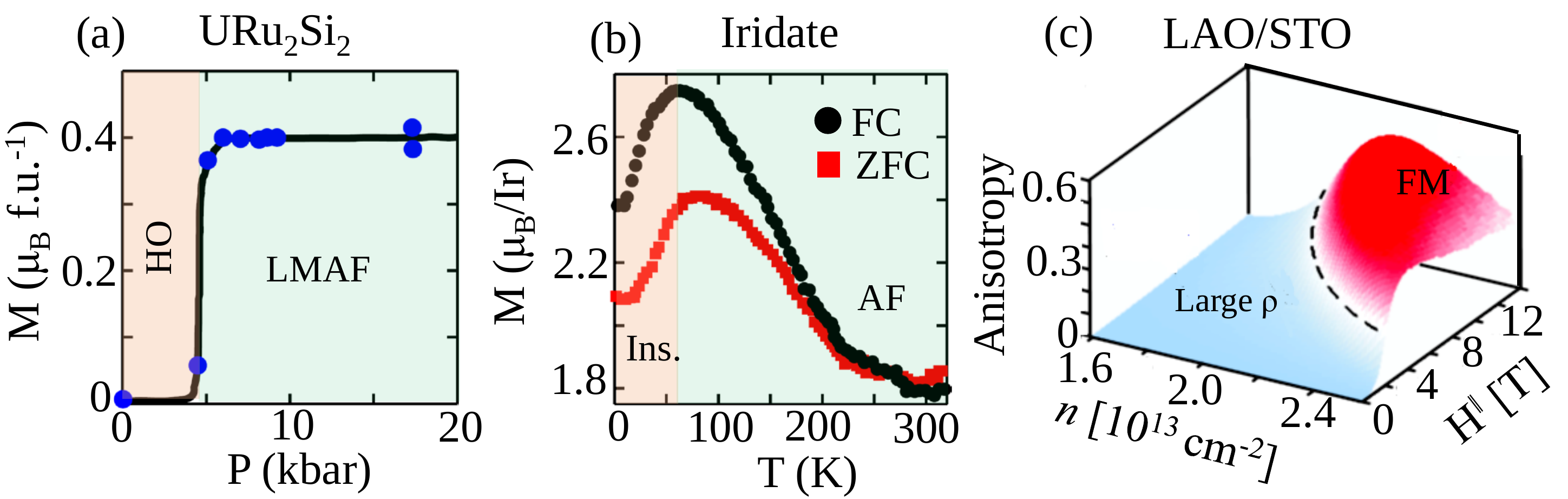}}
\caption{(a) In-plane magnetic moment per U-atom as a function of pressure. The main point of this plot is to emphasize that the magnetic moment of the high-pressure LMAF phase is strongly reduced in the HO state. A phenomenological explanation for this transition from SDW to SODW with decreasing pressure is given in the text. Black line is to guide the eyes. Figure reproduced from Ref.~\cite{PTMoment}. (b) Similarity of the reduction of the magnetic moment with reduced temperature is observed in another SOC iridate. Interestingly, below a characteristic temperature where magnetic moment begins to reduce, the resistivity rises exponentially (metal to insulator transition). Figure reproduced from Ref.~\cite{Iridates}. (c) A very recent observation of a characteristic reduction of preexisting ferromagnetic order below a critical field, carrier density and temperature, at which an `unidentified' phase with large resistivity is reported in the interface of LAO/STO. Figure is reproduced from Ref.~\cite{LAOSTO}. }
\label{fig10}
\end{figure}

Finally we discuss the possible evidence of the SODW in other systems. We recall that SODW quenches any preexisting magnetic moment. This is evident in the {\urusi} phase diagram as a function of pressure, see Fig.~\ref{fig10}(a). Pressure increases the noninteracting bandwidth ($W$), and thus the ratio of $\lambda/W$ decreases, and the possibility of the existence of SODW is reduced while that of the SDW increases, as illustrated in Fig.~\ref{fig4}.

A very similar behavior of the magnetic moment is observed as a function of $T$ for a typical bilayer iridate Sr$_3$Ir$_2$O$_7$\cite{Iridates} and in the LAO/STO interface\cite{LAOSTO} as shown in Figs.~\ref{fig10}(b) and (c), respectively. Iridates exhibit metal-insulator transition (MIT) promoted by strong SOC of 5$d$ orbitals,\cite{BJKim} but no consensus on the origin of MIT has emerged yet. For this given system, an antiferromagnetic (AF) order sets in below 280~K for both field cooled (FC) and zero field cooled (ZFC) samples, however, the resistivity shows a power-law temperature dependence below $T_{AF}$ (a behavior expected for a semi-metallic phase).\cite{Iridates} The resistivity exhibits an abrupt exponential rise which marks the onset of the MIT below $T^*\sim$~70~K, and simultaneously the magnetic moment begins to drop. The new experimental data in the LAO/STO interface which has strong Rashba-type SOC, shows that below a critical magnetic field and carrier density, the preexisting ferromagnetic order is quenched, while the resistivity increases, signaling the onset of a new non-magnetic order parameter, as shown in Fig.~\ref{fig10}(c).\cite{LAOSTO} Based on the similarity of the phase diagram of magnetic moment of {\urusi}, Sr$_3$Ir$_2$O$_7$, and LAO/STO, we envisage that a SODW phase is also present in these materials.\cite{DasIridates}

\begin{figure}[here]
\centering
\rotatebox[origin=c]{0}{\includegraphics[width=.8\columnwidth]{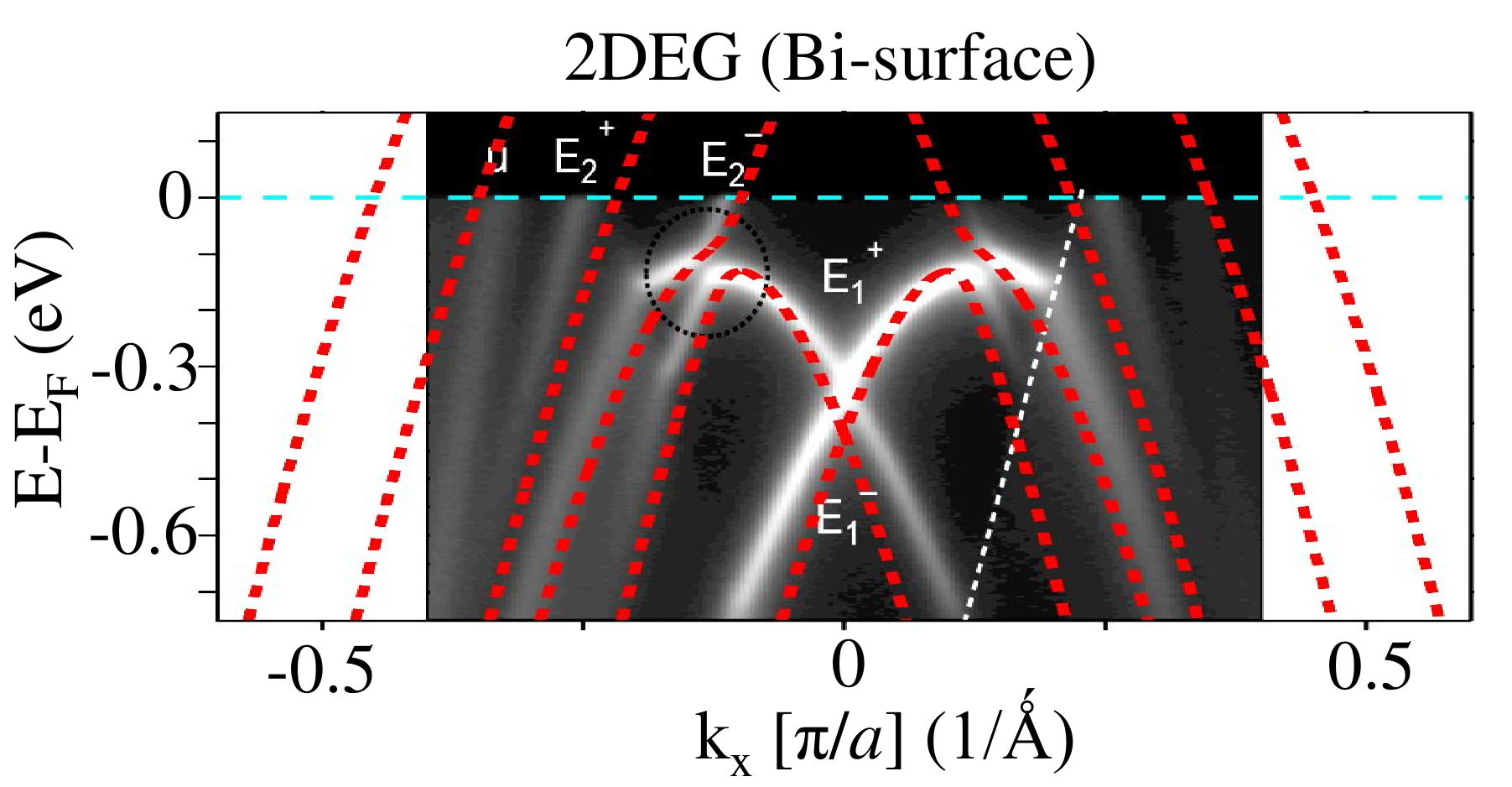}}
\caption{ARPES data of 2 monolayers of BiAg$_2$ thin film plotted in black to while gradient colormap. The red dashed lines are the quasiparticle dispersion in the SODW state in the presence of Rashba-type SOC. Figure reproduced from Ref.~\cite{Das2DEG}.
 }
\label{fig11}
\end{figure}

Another motivation for the search of SODW in 2DEG is the ARPES evidence of a large quasiparticle gapping on the thin film of BiAg$_2$ even when the spin-degenerate point of the Rashba-split bands remain ungapped.\cite{ARPES2DEG} To study the emergence of SODW in this case, we consider a system of two-component Fermi gas in the presence of Rashba-type SOC, which can be represented in the two component fermion field $\psi_{\bm k}=[\psi_{{\bm k},\uparrow}, \psi_{{\bm k},\downarrow}]^{T}$ as $H_0({\bm k}) = \psi_{\bm k}^{\dag}[\xi_{\bm k}{\bf 1} -i\alpha_R({\sigma}_x\sin{k_y}-{\sigma}_y\sin{k_x})]\psi_{\bm k}$. Here  $\xi_{\bm k}$ is the free-fermion dispersion term, modeled by nearest-neighbor electronic hopping $t$ as  $\xi_{\bm k}=-2t[\cos{k_x}+\cos{k_y}]-E_F$, where $E_F$ is the chemical potential. The second term is the 2D lattice generalization of the standard Rashba SOC term $H_{so}=-i\alpha_R(\hat{\bm \sigma}\times\hat{\bm k})_z$, with $\hat{\bm \sigma}$ being the Pauli matrices and $\alpha_R$ being the SOC strength. The helical dispersion spectrum of Hamiltonian $H_0$ is $E_{{\bm k},0}^{\pm}=\xi_{\bm k}\pm\alpha_R[(\sin{k_x})^2+(\sin{k_y})^2]^{1/2}$. The corresponding FS exhibits paramount nesting instability at ${\bf Q}=0.115(\pi,\pi)$ as shown in Ref.~\cite{Das2DEG}.

In this translational symmetry broking case, we define the the Nambu-Gor'kov spinor as $\Psi_{\bm k}=[\psi_{{\bm k},\uparrow}, \psi_{{\bm k},\downarrow}, \psi_{{\bm k}+{\bm Q}_i,\uparrow}, \psi_{{\bm k}+{\bm Q}_i,\downarrow}, ...]^T$, where $i=1-4$ corresponds to four-different ${\bm Q}$-vectors. In this basis, the auxiliary SODW field $\Delta ({\bm k})=V\psi_{{\bm k}+{\bm Q},\nu}^{\dag}[\sigma_z\otimes\sigma_x]_{\nu\nu^{\prime}}\psi_{{\bm k},\nu^{\prime}}$, where $V$ is the contact potential. Correspondingly, the inverse single-particle Green's function in the fermionic Matsubara frequency is obtained as
\begin{eqnarray}
G^{-1}=\left(
\begin{array}{ccc}
i\omega_n{\bf 1}-H_0({\bm k}) & {\bm \sigma}_x\Delta&\ldots\\
{\bm \sigma}_x\Delta^{\dag} & i\omega_n{\bf 1}-H_0({\bm k}+{\bm Q}_i)&\ldots\\
\vdots&\vdots&\ddots
\end{array}
\right).
\label{greenfn}
\end{eqnarray}
The other terms belong to the three values of ${\bm Q}_i$. At the mean-field level, $\Delta({\bm k})$ represents the gap parameter which is same for all values of ${\bm Q}$. As in the case of the Hamiltonian for {\urusi} in Eq.~\ref{Hint}, the above Hamiltonian also has resemblance with the Hamiltonian for the topological insulator in heterostructure of Rashba-type SOC 2DEG proposed in Ref.~\cite{DasTI} in which the SODW term is replaced by the interlayer tunneling amplitude. Taking the trace of the above Green's function, we obtain the excitation spectrum as
\begin{eqnarray}
&E_{\bm k}^{\mu,\nu}=S_{+}^{\nu}+\mu [(S_{-}^{\nu})^2+\Delta^2]^{1/2},\nonumber\\
&{\rm with}\hspace{0.15cm}S^{\nu}_{\pm}({\bm k})=(E_{{\bm k},0}^{\nu}\pm E_{{\bm k},0}^{\bar{\nu}})/2,
\label{eq:eig}
\end{eqnarray}
where $\bar{\nu}=-\nu=\pm$ is the helical index and $\mu=\pm$ is the split band index due to translational symmetry breaking. For a self-consistent value of $V\sim$310~meV, we obtain a quasiparticle gap of $\Delta=$120~meV, which agrees well with the gap amplitude measure for 2 monolayers of BiAg$_2$.\cite{ARPES2DEG}. Experimentally it was also demonstrated that the quasiparticle gap is tunable by changing the number of layers of BiAg$_2$ layer which we modeled by changing interaction parameter $V$, while all other parameters including Rashba SOC remain constant. We plot the corresponding calculated band dispersion on top of the experimental dispersion in Fig.~\ref{fig11}, and find good correspondence between the theory and experiment.

In both experiment and theory, two key features should be noted here. We notice the presence of shadow bands (or reconstructed bands due to translational symmetry breaking) which convey the fact that the quasiparticle gap is not a single-particle gap. Furthermore, since the Kramer's degeneracy at the $\Gamma$-momentum point remains ungapped, this particular phase is not a SDW. Such a reconstruction of FS and FS nesting has also been documented in a quantum wire of atomic Pb grown on Si(557) having strong SOC spitting.\cite{Quantumwire} Our SODW model for the Rashba-type SOC which is a generic feature arising in many interfaces and surface states (sometimes in bulk states as well) establishes that the SODW phase may be present in many systems other than {\urusi}.

\section{Topological properties of the SODW}
The similarity of the Hamiltonian for the SODW in {\urusi} in Eq.~\ref{Hint} and Rashba-type SODW in Eq.~\ref{greenfn} with that of the non-interacting TI in Refs.~\cite{SCZNP,DasTI} motivates us to evaluate the topological invariant index of the interacting Hamiltonian in Eq.~\ref{Hint}. To characterize the topological phenomena, we revisit the Fu-Kane classification scheme\cite{FuKane} which implies that if a time-reversal invariant system possess an odd value of $Z_2$ invariant index, then the system is guaranteed to be topologically non-trivial. $Z_2$ index is evaluated by the time-reversal invariant index $\nu_i=\pm 1$, if defined, for all filled bands as $Z_2=\nu_1\nu_2 ... \nu_n$, where $n$ is the total number of orbitals in the Fermi sea. A more efficient method of determining the topological phase is called the adiabatic transformation scheme used earlier in realizing a large class of topological systems, especially when $Z_2$ calculation is difficult.\cite{DasTI} In this method, the non-trivial topological phase of a system can be realized by comparing its band-progression with respect to an equivalent trivial topological system. {\urusi} is topologically trivial above the HO state, i.e. $Z_2^0=+1$. The gap opening makes the top of the valence band (odd parity) to drop below $E_F$ as shown in Fig.~\ref{fig6}(b). Thereby, an odd parity gained in the occupied level endows the system to a non-trivial topological metal. To see that we evaluate the topological index for the HO term as $\nu_{ho} = \int d{\bm k} \Omega({\bm k})$, where the corresponding Berry curvature can be written in terms of ${\bm d}$-vector as $\Omega=\hat{d}\cdot\left(\frac{\partial \hat{d}}{\partial k_x}\times\frac{\partial \hat{d}}{\partial k_y}\right)$ with $\hat{d}={\bm d}/|{\bm d}|$. The components of the ${\bm d}$ vector is obtained from the SODW order parameter given in Eq.~(\ref{SODW}). When the ${\bm d}$ vector has a odd parity symmetry, it is easy to show that $\nu_{ho}=-1$ which makes the total $Z_2$ value of the HO phase to be $Z_2=Z_2^0\times\nu_{ho}=-1$, and hence we prove that the HO gap is a topologically non-trivial phase. The consequence of a topological bulk gap is the presence of surface states.\cite{FuKane} In our present model, we expect two surface states of opposite spin connecting different orbitals inside the HO gap. As the system is a weak topological semimetal, the surface states are unlikely to be topologically protected. The spin-orbit locking of these states can be probed by ARPES using circular polarized incident photon which will be a crucial test of this postulate. Furthermore, the topological nature of the HO state can lead to non-trivial ground state in the LMAF and SC state, if they uniformly coexist. For example, topological HO with LMAF phase may host non-Abelian axion particle, or in the SC state may give rise to a topological SC.\cite{Goswami}

\section{Discussion and conclusion}

To conclude, in this article we presented detailed analytical calculation and numerical results of SODW as a candidate proposal of the long-sought HO phase of the heavy-fermion metal {\urusi}. The proposal of the SODW is motivated by the `magnetic dichotomy' phenomena at this phase transition as well as the FS instability between two spin-orbit split bands as deduced consistently by DFT calculation and ARPES results. DFT calculation demonstrated the presence of a logarithmic-like divergent susceptibility at ${\bf Q}=0.5(\pi,\pi,0)$ in the BCT crystal structure which connects inter-orbital Fermi momenta. As a result, a density wave originates which involves both spin and orbital flips at the same time between two in-plane sublattices of U atoms which is thus termed as SODW. The corresponding SODW gap opens in the electronic and magnetic spectrum, but no static magnetic moment is induced due to time-reversal invariance imposed by SOC.

We formulated a general two orbital model Hamiltonian including SOC. The similarity of the SODW Hamiltonian (Eq.~\ref{Hint}) with that of the three-dimensional topological insulator\cite{SCZNP} is noted. The numerical results were deduced by including DFT bandstructure for the noninteracting bands. We found that FS gapping occurs along the zone diagonal direction in the BCT crystal, in agreement with ARPES data.\cite{Durakiewicz} We also computed the spin-excitation dispersion which showed a gapped upward dispersion at the incommensurate wavevector, in agreement with INS data.\cite{INS} Finally, we showed that the SODW order yields a second resonance mode at ${\bf q}\sim$0, which can be probed by polarized INS or ESR experiment at zero magnetic field. Most importantly, despite dynamical magnetic signal, the static magnetic moment was shown analytically to be zero.

The feedback effect of the time-reversal invariance can be tested by applying magnetic field to the HO state. Due to symmetry reason, the HO can be destroyed by magnetic field, but to overcome the finite gap opening, a finite magnetic field is required to compensate for the HO gap at zero temperature. For this reason, SODW predicts the presence of a QCP along the field axis. With a rough estimation of the critical field (assuming the Zeeman energy to be equal to the gap energy), we found a good agreement with experiment.\cite{HarrisonQCP}

We also gave evidence of the presence of SODW in other SOC systems such as iridates, BiAg$_2$ thin film, and LAO/STO interface. For BiAg$_2$ thin film, ARPES exhibits the presence of unusual quasiparticle gapping and shadow bands while the spin-degenerate point at $\Gamma$-momentum is ungapped.\cite{ARPES2DEG} These results are not reproduced by corresponding DFT band structure. We extended the SODW concept to the Rashba-type SOC bandstructure and explained the nature of this unusual gapping phenomena.\cite{Das2DEG} Similarly, in iridates\cite{Iridates} and LAO/STO interface,\cite{LAOSTO} we found that a preexisting magnetic moment is quenched below a characteristic temperature and magnetic field, respectively, indicating the presence of an unusual phase. More interestingly, for iridates, it was also shown that a metal-insulator transition occurs at the same temperature where the magnetic moment decreases, suggesting that magnetic ordering is unlikely to be responsible for the MIT behavior in this compound. Due to the presence of SOC and the possible presence of FS nesting, we envisioned to explain these unusual properties by SODW. However, no theoretical calculation for iridates and LAO/STO were presented in this article.

Taken together, SODW was shown to be a new state of matter which breaks translational symmetry, but not time-reversal symmetry. This is why SODW is fundamentally distinct from an inter-orbital SDW. Also the contributing interaction terms for SODW are inter-orbital interaction $V$ and pair interaction $J^{\prime}$, as given in Eq.~(\ref{sceqn}), while for the inter-orbital SDW the pair scattering is replaced by the Hund's coupling $J$. In the present SODW model both contributing orbitals are itinerant and they are split by SOC.

Finally, we comment on two future research directions to further explore the nature of SODW in {\urusi}. (1) The competition between HO and LMAF phase as a function of pressure has remained unexplained. As shown in the phase diagram in Fig.~\ref{fig9}, the HO phase at ambient pressure goes through a first order phase transition around 0.5~GPa, and at higher pressure it coexists and/ or competes with the LMAF phase. With pressure, however, the HO transition temperature increases, rather than decreases. To explain this phenomenon, our strategy is to study the competition between SDW and SODW. We expect that as pressure increases, the effective non-interacting bandwidth increases. Therefore, $\lambda/W$ decreases which, according to Fig.~\ref{fig4} weakens SODW, and favors SDW. The corresponding Ginzburg-Landau criterion for this problem is deduced in Sec.~4.2. (2) As magnetic field is expected to reduce the SODW order for all systems due to time-reversal invariance, one would expect to gain more quasiparticle state on the FS with increasing field at a fixed temperature. At $T=0$ the resistivity of a system is determined by the quasiparticle scattering on the FS without the inclusion of scattering lifetime or the imaginary part of the self-energy, Therefore, with decreasing field, one would expect a larger FS gapping and thus larger resistivity. This is seen in LAO/STO interface\cite{LAOSTO} and in iridates,\cite{Iridates}; and we expect to see similar effect in {\urusi}.

\section*{Acknowledgement(s)}
The author acknowledges valuable discussion with A. Leggett, J. Mydosh, P. Coleman, M. J. Graf, P. Wo\"elfle, J.-X. Zhu, J. Haraldsen, P. Riseborough, A. F. Santander-Syro. The author also expresses gratitude to A. V. Chubukov and A. Rahmanian for discussion of SODW formalism in iridates. The work is supported by the U.S. DOE through the Office of Science (BES) and the LDRD Program and faciliated by NERSC computing allocation.

\end{document}